\documentclass[twoside,12pt]{article}

\pdfoutput=1

\usepackage{epsfig,graphicx}

\begin{document}

\title{Phases of QCD: \\
Summary of the Rutgers Long Range Plan Town Meeting, January 12-14, 2007
}

\author{
Peter Jacobs\\
{\it Lawrence Berkeley National Laboratory}\\
\\
Dmitri Kharzeev\\
{\it Brookhaven National Laboratory}\\
\\
Berndt M\"{u}ller\\
{\it Duke University}\\
\\
Jamie Nagle\\
{\it University of Colorado}\\
\\
Krishna Rajagopal\\
{\it MIT}\\
\\
Steve Vigdor\\
{\it Indiana University}\\
}

\maketitle


\newpage

\tableofcontents

\newpage


\def\ET{\mbox{$E_T$}}
\def\pT{\mbox{$p_T$}}
\def\Qsq{\mbox{$Q^2$}}

\def\qhat{\mbox{$\hat{q}$}}

\def\RAA{\mbox{$R_{AA}$}}

\def\pizero{\mbox{$\pi^0$}}
\def\JPsi{\mbox{$\mathrm{J/}\psi$}}
\def\PsiPr{\mbox{$\psi\prime$}}
\def\Chic{\mbox{$\chi_c$}}
\def\UpsOneS{\mbox{$\Upsilon(1S)$}}
\def\UpsThreeS{\mbox{$\Upsilon(3S)$}}
\section{Executive Summary}
\label{ExecSummary}

This White Paper summarizes the outcome of the Town Meeting on Phases
of QCD that took place January 12-14, 2007 at Rutgers University, as
part of the NSAC 2007 Long Range Planning process. The meeting was
held in conjunction with the Town Meeting on Hadron Structure,
including a full day of joint plenary sessions of the two
meetings. Appendix \ref{Program} contains the meeting agenda.

This Executive Summary presents the prioritized recommendations that
were determined at the meeting. Subsequent chapters present the
essential background to the recommendations. While this White Paper is
not a scholarly article and contains few references, it is intended to
provide the non-expert reader with a complete and nuanced case
supporting the recommendations.

The prioritized recommendations of the Phases of QCD community are as
follows:

\begin{enumerate}

\item Our central goal is a dramatic advance in our 
understanding of QCD Matter, through quantitative comparison of theory
and experiment to determine the properties of the strongly interacting
Quark-Gluon Plasma discovered in the initial phase of RHIC operations,
and through further exploration of the QCD phase diagram at non-zero
baryon density where a critical point has been predicted. The
essential requirements for the success of this scientific program are
therefore our highest priorities:

\begin{itemize}

\item Effective utilization of the RHIC facility and 
completion of the ongoing detector upgrade program; 

\item The RHIC II luminosity upgrade, which will 
enable quantitative study of key rare processes;

\item Strong support for the ongoing theoretical studies 
of QCD matter, including finite temperature and finite baryon density
lattice QCD studies and phenomenological modeling, and an increase of
funding to support new initiatives enabled by experimental and
theoretical breakthroughs.

\end{itemize}

\item We strongly recommend significant and timely 
participation of U.S. groups in the LHC heavy ion program, which will
study QCD matter at the highest energy densities and temperatures
available in the laboratory. This program will test and extend the
insights reached in the RHIC program, and has the potential to make
important new discoveries about QCD Matter.

\item An Electron-Ion Collider (EIC) facility is the highest
priority of the QCD community for new construction after the JLab 12
GeV and the RHIC II luminosity upgrades. EIC will address compelling
physics questions essential for understanding the fundamental
structure of matter: 

\begin{itemize}

\item Precision imaging of sea-quarks and gluons to
determine the full spin, flavor and spatial structure of the nucleon;

\item Definitive study of the universal nature of strong gluon
fields manifest in nuclei.  

\end{itemize}

This goal requires that R\&D resources be allocated for expeditious
development of collider and experimental design.

\item Nuclear theorists play an essential role in the development of 
future research directions, the interpretation of experiments, and the 
articulation of their impact to the broader physics community. In 
many cases, significant contributions originate from our young 
scholars. In addition to the continuing success of this sustained 
effort, a number of key theoretical challenges remain. Meeting
these challenges through targeted new investments is critical to
realizing the full impact of the scientfic program outlined in this 
Long Range Plan. 

\begin{itemize}
\item We strongly recommend new investments in the next generation of 
nuclear theorists who are critical to the future of the field, and 
targeted support for initiatives to solve the key scientific problems 
identified in this LRP
\end{itemize}

\item Education and Outreach are of crucial importance to the Nuclear 
Physics community and to the nation as a whole. We strongly support
the efforts of NSAC, the funding agencies, and other bodies, to expand
education and outreach activities in Nuclear Physics at all levels,
from elementary school through graduate education, and to help ensure
a scientifically literate citizenry.

\item We support the creation of a coordinated national
program in Accelerator Science and Technology, including a PI-driven
program targeted at technologies that will enable major advances in
Nuclear Physics.

\end{enumerate}


\section{Phases of QCD: Current Status}
\label{HIStatus}
The RHIC accelerator complex and its complement of detectors have
exceeded their initial scientific promise. When the RHIC physics
program was planned, the hope was that collisions of heavy nuclei at
energies up to 200 GeV per nucleon pair would result in the formation
of a new type of strongly interacting matter, the quark-gluon
plasma. A further hope was that the collisions would produce
sufficient evidence of the nature of this matter to explore its
physical properties and address the fundamental question: What is the
structure of matter at the highest energy densities?  Additionally,
some speculated that the matter would be weakly coupled
(describable perturbatively) and that indications of a strong first
order phase transition might be observed.  As a result of the
experiments at RHIC, we now know that the matter is far from weakly
coupled and a strong first-order phase transition can be excluded.
Perhaps even more exciting, we know for certain that a new type of
strongly interacting thermalized matter is, indeed, produced in nuclear
collisions, and we have begun quantitative measurements of its
structure and properties.

In the first six runs (2000-2006), BRAHMS, PHENIX, PHOBOS and STAR
have collected data from Au+Au, d+Au, Cu+Cu, and p+p collisions, with
Au+Au collisions having been studied at four collision energies
($\sqrt{s_{\rm NN}} =$ 19.6, 63, 130 and 200 GeV). The largest data
samples were collected at the highest energy of $\sqrt{s_{\rm NN}} =$
200 GeV.  The ability to study proton-proton, deuteron-nucleus, and
nucleus-nucleus collisions with identical center-of-mass energies at
the same facility has been the key to systematic control of the
measurements. Nearly all observables have been studied as a function
of collision centrality and of the emission angle relative to the
reaction plane, thereby providing complete control over the collision
geometry. It is noteworthy that the results obtained by the four RHIC
experiments are overall in excellent quantitative agreement.

\subsection{The RHIC Discoveries}

Results from the first five years of RHIC operations with heavy ions
have provided evidence for the creation of a new state of thermalized
matter at unprecedented energy densities (more than 100 times larger
than that of normal, cold nuclear matter) which appears to exhibit
almost perfect hydrodynamical collective behavior.  Among this
evidence, four fundamental new discoveries stand out:
\begin{itemize}

\item {\bf Near-Perfect Liquid}:  The measured hadron spectra and 
their angular distributions bear witness to the enormous collective
motion of the medium.  In addition, measurements of non-photonic
electrons, attributed to the decays of open charm hadrons, indicate 
that even heavy quarks flow with the bulk medium.
These observations are in agreement with the hydrodynamic behavior of
a nearly inviscid, i.~e.\ viscosity-free, liquid -- often characterized
as a ``perfect liquid'' -- and point to a rapid thermalization and
equilibration of the matter.

\item
{\bf Jet Quenching}: The strong quenching of jets, observed in central 
Au+Au collisions via the suppression of particle production at high 
transverse momentum and the dramatic modification of jet correlations,
are evidence of the extreme energy loss of partons traversing matter
containing a large density of color charges.

\item {\bf Novel Hadronization}: The large, unexpected enhancement of 
baryon and anti-baryon production, relative to meson production, at 
intermediate transverse momentum, 
together with the observed scaling of the collective motion of hadrons
with the number of valence quarks, suggests that hadrons form by
parton recombination after the collective flow pattern is established.

\item{\bf Novel phenomena at high parton density}: 
The RHIC experiments have observed low multiplicity of produced
particles, compared to most expectations, together with a suppression
in the production of high-transverse momentum particles at forward
rapidity in deuteron-gold interactions. These phenomena may be the
first indications of parton saturation inside the colliding nuclei.

\end{itemize}

\noindent
These discoveries stand out, but many other results have been obtained
which contribute important facets to the overall picture of the formation 
of an equilibrated QCD medium of unprecedented energy density
and endowed with novel and unexpected properties. 

The RHIC experiments confirmed with high statistics and often better
systematics important features of ultra-relativistic heavy-ion
collisions that were previously discovered at lower energies. For
example, all hadron abundance ratios are characterized by a chemical
equilibrium distribution with chemical freeze-out temperature $T_{\rm
ch} = 160-170$ MeV, a value that is observed to be independent of the
collision system and the collision centrality. First results on
charmonium production reveal a striking similarity of suppression in
the medium to results at much lower beam energy, contrary to many
expectations.  Direct photon production at high transverse momentum
has been measured and is in excellent agreement with pQCD
expectations.  Direct photon measurements at lower transverse momentum
where significant thermal radiation contributions may be seen are
underway.

\subsection{Detailed Discussion: Experiment and Theory}

We now present these results and their implications in more detail.

\subsubsection{Near-Perfect Liquid} 

A ``perfect'' liquid is one that obeys the equations of ideal 
hydrodynamics without shear or bulk viscosity. In practice, any liquid 
must have a nonvanishing viscosity, because the mean 
free path of thermal excitations cannot be zero. Quite generally, a low shear
viscosity implies a large transport cross section and thus strong coupling.
One of the exciting theoretical discoveries of the past few years is the 
insight that there may exist a lower bound on the dimensionless ratio 
between the shear viscosity and entropy density of any fluid 
($\eta/s \geq 1/4\pi$, see Section 2.3.3). Thus, in reality, a perfect liquid 
is a fluid that attains this lower bound. As we discuss in the following, 
measurements of the hadronic collective flow indicate that the matter 
produced at RHIC is, indeed, not far from this bound on $\eta/s$.

\begin{figure}[tbp!]
\centering
\includegraphics[width=0.9\textwidth]{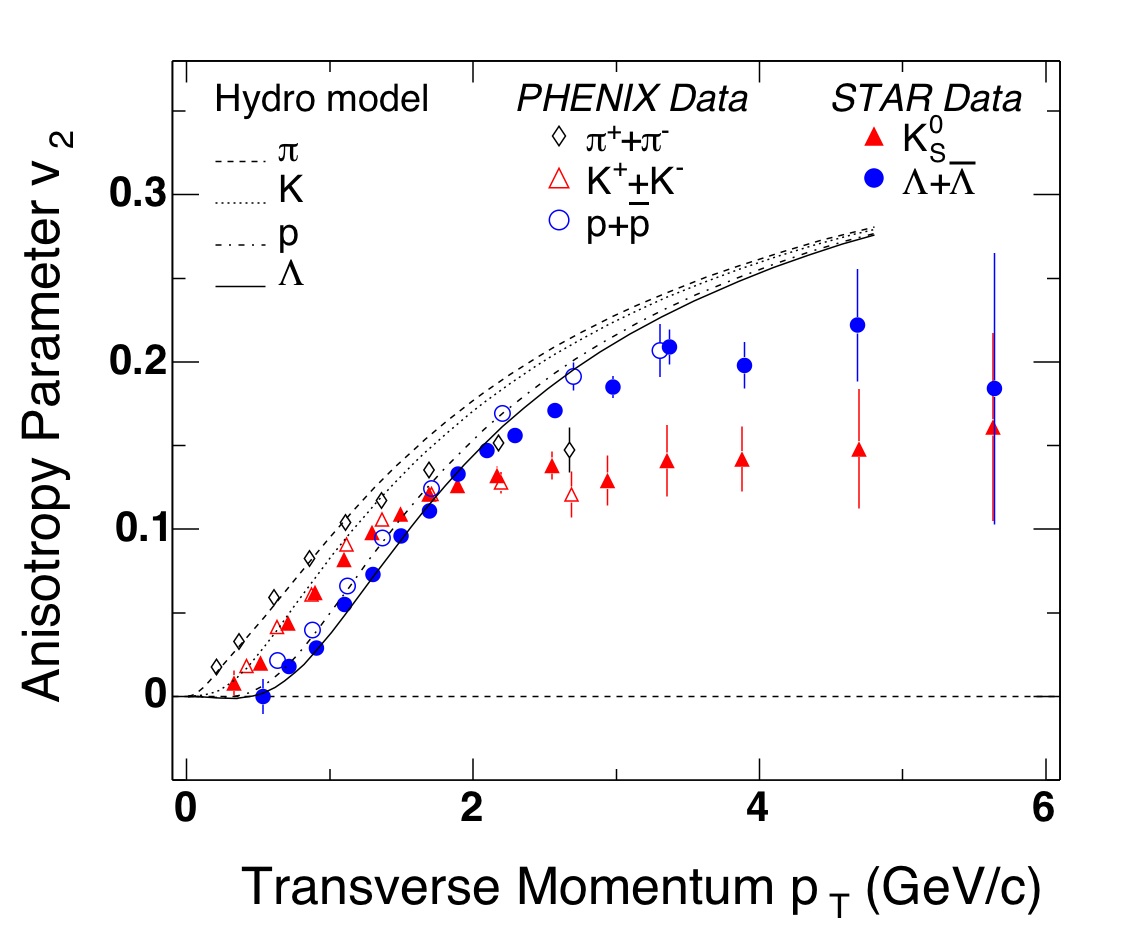}
\caption{Compilation of STAR and PHENIX data on elliptic flow $v_2$ for identified hadrons,
plotted as a function of transverse momentum \pT\ and compared with
hydrodynamic predictions \cite{V2Compilation}. The elliptic flow is a measure of the
anisotropic pressure-driven expansion in off-center collisions.  Note
that the bulk of the particle production occurs at less than 2
GeV/c. Collective motion of hadrons is expected to disappear above
$\pT > 1.5-2$ GeV/c.
\label{fig2.1}}
\end{figure}

The abundances of the produced hadrons at midrapidity with transverse
momenta below about 2 GeV/c, the shapes of their transverse momentum
spectra, and the elliptic flow of these hadrons can be very well
described by relativistic hydrodynamics for a perfect liquid with an
equation of state similar to the one predicted by lattice QCD
(Fig.~\ref{fig2.1}). While strong collective flow had been observed
previously in lower energy heavy-ion collisions, hydrodynamic models
were never before able to provide an equally successful quantitative
description of the data. The best overall description of the RHIC data
is obtained if the ideal hydrodynamical evolution of a quark-gluon
plasma during the early expansion stage is combined with a realistic
hadronic cascade after hadronization, and if an equation of state like
that obtained from lattice QCD is employed. To reproduce the magnitude
of the observed radial and elliptic flow it is necessary to assume
that the produced matter thermalizes very quickly, on a time scale of
less than 1 fm/c, and builds up thermodynamic pressure whose gradients
drive the collective expansion.

Hydrodynamic calculations that reproduce the experimental data
indicate that at thermalization time the energy densities must be at
least 15 GeV/fm$^3$, i.~e.\ 15 times the energy density needed for
color deconfinement. In fact, even if one
applies only the principle of energy conservation to the measured
produced transverse energy in the collision, neglecting any energy
lost to longitudinal work during the expansion, and uses any
reasonable estimate for the initial volume of the fireball at
thermalization, one also obtains a lower limit for the initial energy
density which is about an order of magnitude above the critical value
for deconfinement.

\begin{figure}[tbp!]
\centering
\includegraphics[width=0.7\textwidth]{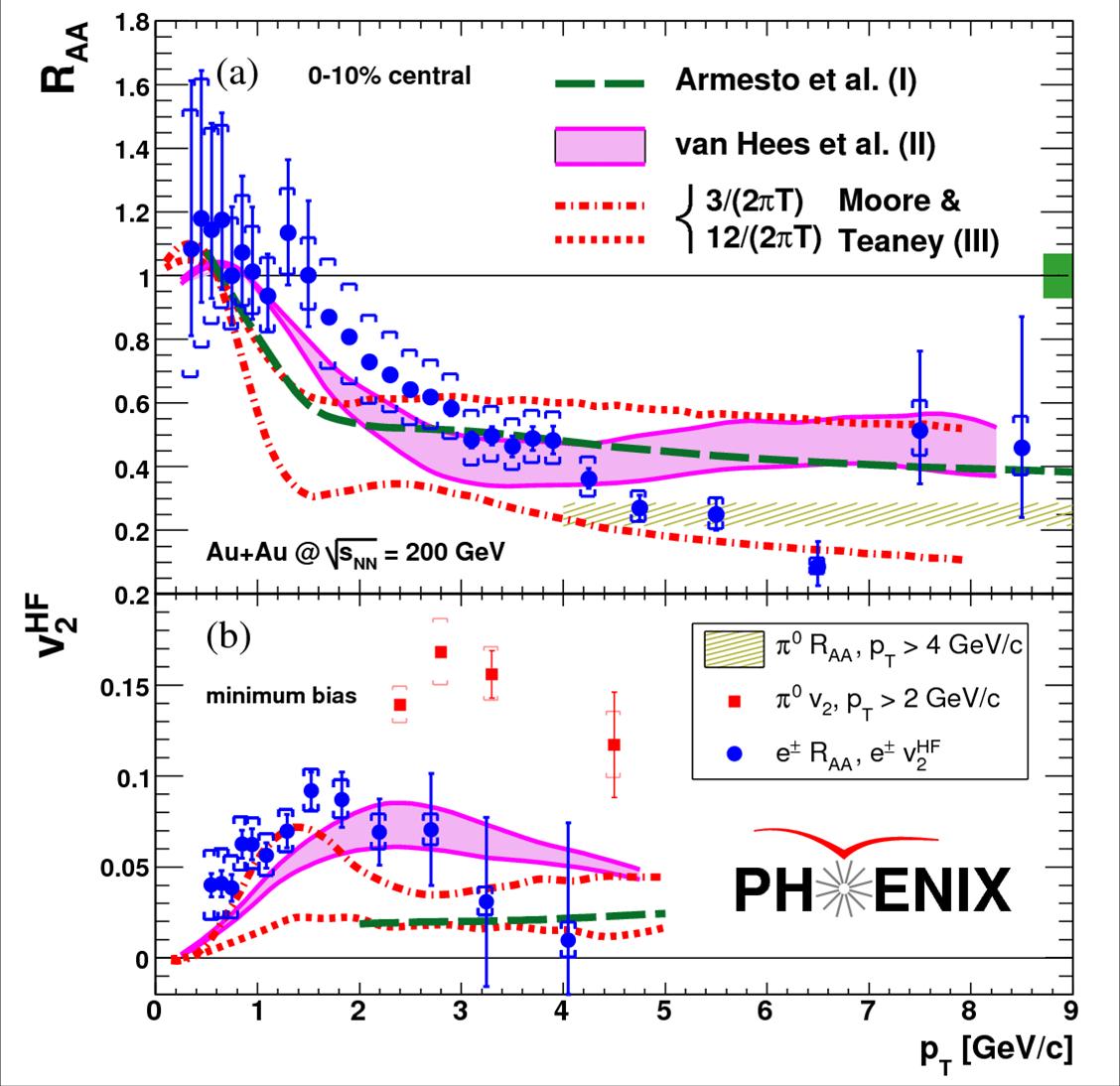}
\caption{Figure from \cite{PhenixHF}. (a) The nuclear modification factor $R_{AA}$ of heavy-flavor 
electrons in 0-10\% central Au+Au collisions compared with \pizero\ 
data and model calculations. The nuclear modification factor is the ratio 
of the cross section per nucleon-nucleon collision measured in a heavy 
ion collision divided by the cross section measured in p+p collisions.  If 
there were no nuclear effects it would be unity. 
(b) $v_2$ of heavy-flavor electrons in minimum bias
collisions compared with \pizero\ data and the same models.
\label{fig2.1b}}. 
\end{figure}

The evidence for fast thermalization, the observation of large elliptic
flow even for multi-strange (anti-) baryons and charmed hadrons, and
the good agreement of ideal hydrodynamical models assuming a vanishing 
shear viscosity of the matter during the early phase of its expansion, indicate 
that the extremely hot and dense medium created in the collision is a 
strongly coupled medium with the properties suggestive of a nearly perfect 
liquid. Its apparently almost complete absence of viscosity contrasts strongly
with intuitive expectations by many scientists in the field that the
quark-gluon plasma would exhibit perturbative, gas-like behavior
characterized by weakened interactions among its partonic
constituents.

The unanticipated success of ideal relativistic hydrodynamics to
describe the collective flow imprinted on the hadron spectrum from
nuclear collisions at RHIC has made it possible to develop a
compelling foundation for the dynamical treatment of almost the entire
collision process (except for the process of thermalization itself),
which can serve as a basis for future, more refined treatments. In
this framework, the densest stage of the collision, in which the
matter is in the quark-gluon plasma phase, is described in terms of
ideal relativistic hydrodynamics. The inputs for this description are
the equation of state of the matter and the initial conditions, given
in terms of the energy density of the matter at the moment of
thermalization. The final, much more dilute hadronic stage of the
collision is described by a Boltzmann cascade of binary hadronic
interactions, which is tracked up to the disintegration of the matter
into individual, free-streaming hadrons. Two independently developed
implementations of this concept have had remarkable success in
describing the global features of the heavy ion reactions (spectra,
flow anisotropies, hadron ratios, etc.).

\newpage

\subsubsection{Jet Quenching} 
\label{JetQuenchingStatus}

\begin{figure}[htbp!]
\centering
\includegraphics[width=0.7\textwidth]{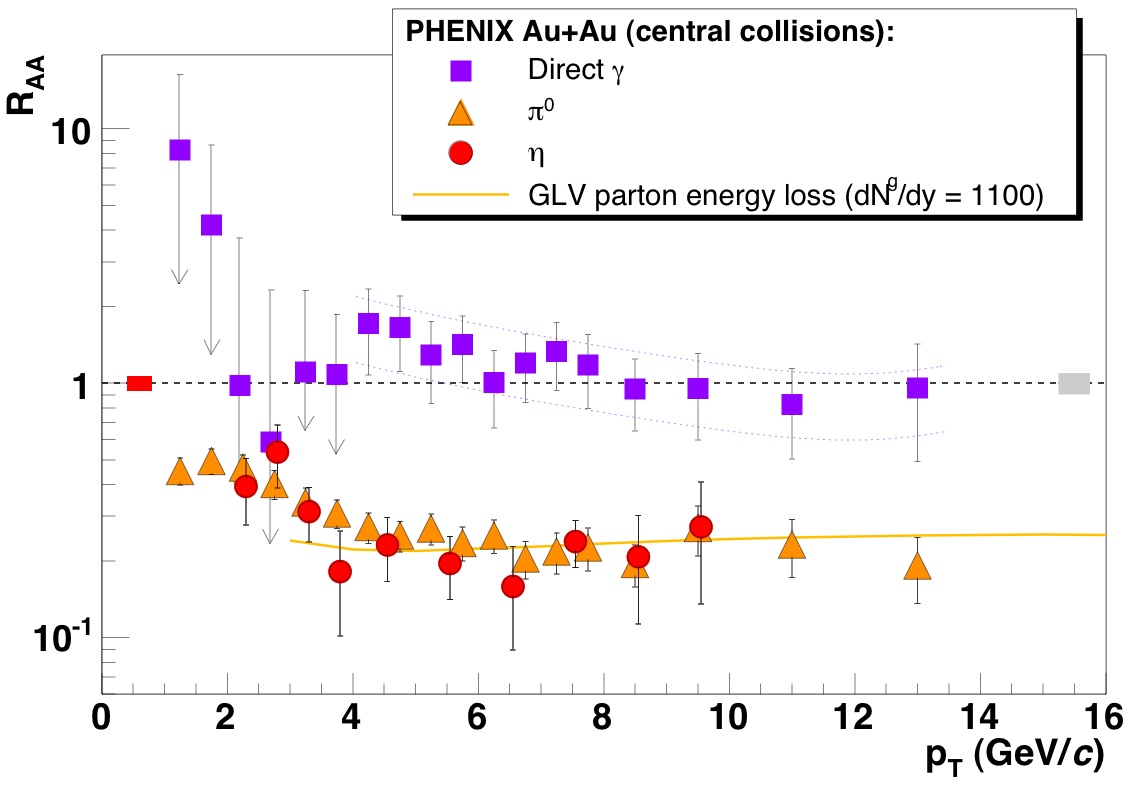}
\caption{Nuclear modification factor $R_{AA}(\pT)$ for photons ($\gamma$), 
$\pizero$ and $\eta$ mesons in central Au+Au collisions \cite{PhenixRAA}. The nuclear
modification factor is the ratio of the cross section per
nucleon-nucleon collision measured in a heavy ion collision divided by
the cross section measured in p+p collisions.  If there were no
nuclear effects it would be unity. Note the strong suppression of the
mesons and the lack of suppression for the photons, which do not
interact with the final state medium.
\label{fig2.2}
}
\end{figure}

The medium created in collisions at RHIC shows evidence of strong
interactions not only among its constituents, but also with hard
colored penetrating probes, such as energetic quarks and gluons created
at the very beginning of the collision and propagating outward through
the reaction zone.  High transverse momentum hadrons, which arise from
the fragmentation of such hard partons, are found to be suppressed in
central Au+Au collisions by a factor of up to five relative to the
experimental proton-proton baseline (when normalized to the number of
pairwise nucleon-nucleon interactions)(Fig.~\ref{fig2.2}). 

In contrast, the production rates and spectra of direct photons, which escape from the
collision without further interaction, agree well with expectations
based on perturbative QCD. Confirmation of this interpretation comes
from two other experimental observations: (1) The strong suppression
of high $p_T$ hadrons is not observed in d+Au collisions, which rules
out initial state effects associated with possible modifications of
the parton distributions in heavy nuclei; (2) when triggering on a
high-$p_T$ hadron with transverse momentum of up to 10 GeV/c, the data
show its opposite partner jet even more strongly quenched in central
Au+Au collisions.

\begin{figure}[tbp!]
\centering
\includegraphics[width=0.49\textwidth]{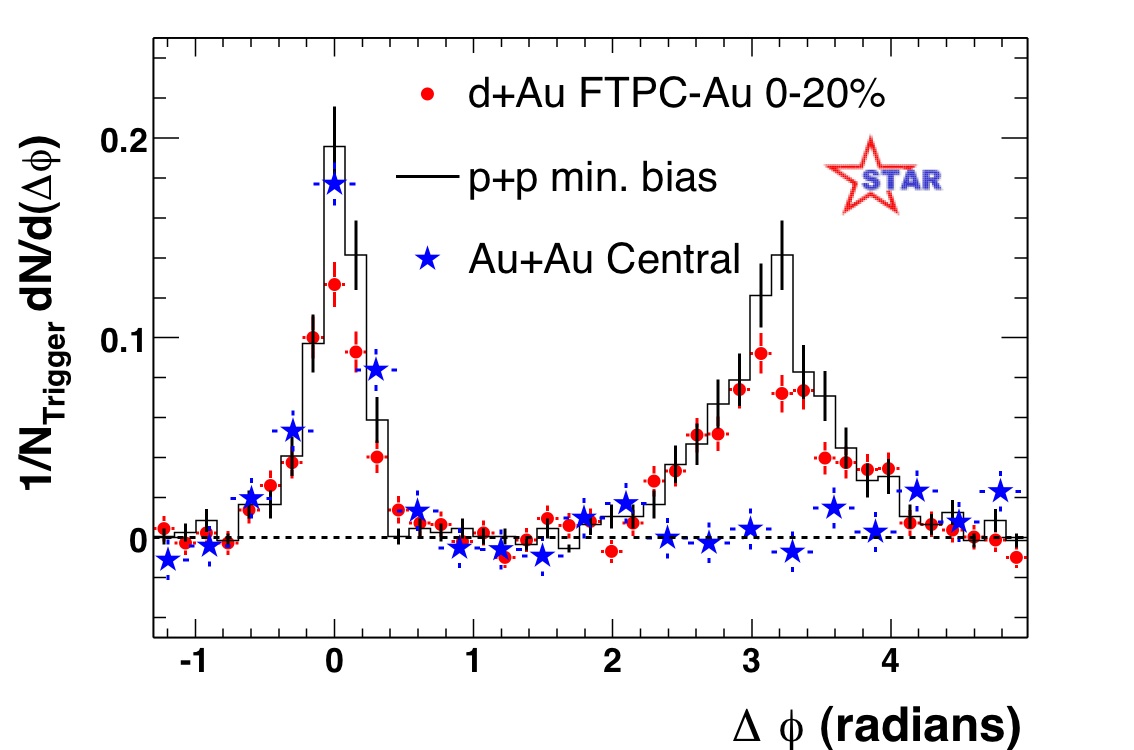}
\includegraphics[width=0.49\textwidth]{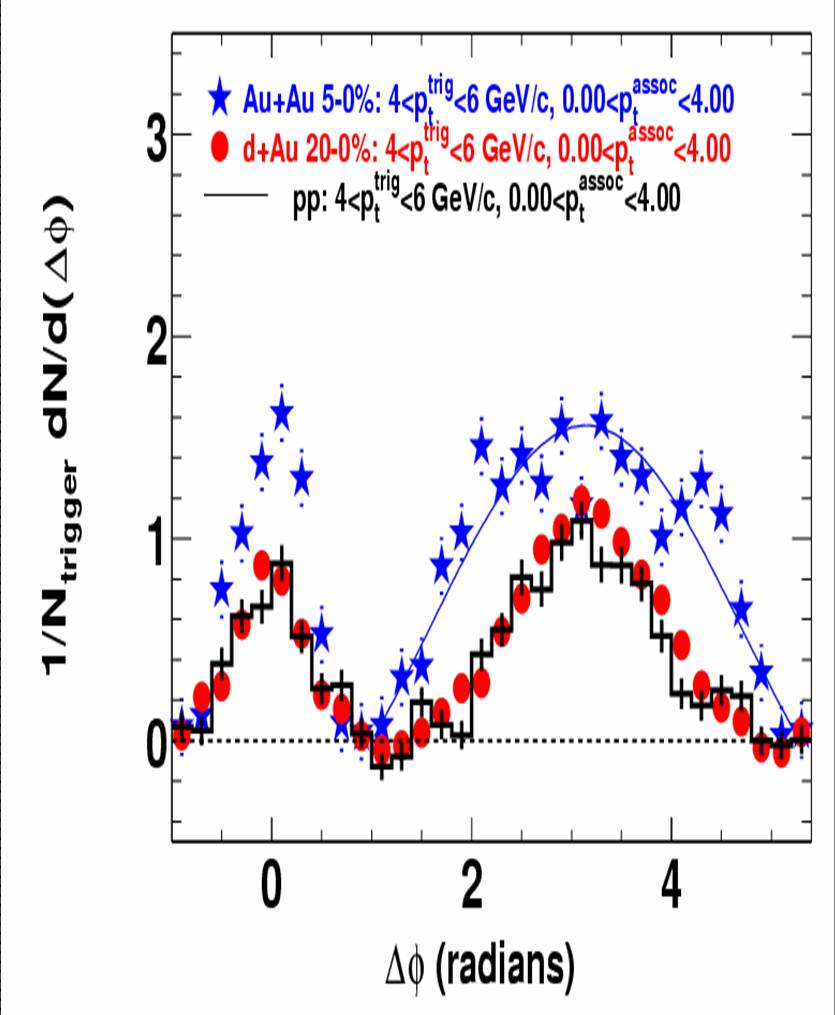}
\caption{Dihadron correlations in azimuthal angle. Left: pairs of high 
momentum hadrons, showing suppression of leading recoil particles in
central Au+Au relative to pp and dAu collisions
\cite{STARDihadronHighPt}. Right: high momentum trigger and low
momentum associated hadron, showing enhanced yield of recoiling soft
jet fragments in Au+Au \cite{STARDihadronLowPt}.
\label{fig2.3}
}
\end{figure}

The large momentum scale associated with the primary jet production
vertex combined with enhanced momentum transfer to the jet on its way
out by the dense medium permit a rigorous formulation of jet quenching
in the framework of perturbative QCD. Although the formalism can be
cast in several different forms, all formulations assign the quenching
power of the medium to a unique transport coefficient, the jet
quenching parameter $\hat{q}$, which measures the transverse momentum
broadening of a hard parton propagating through the medium. The
parameter $\hat{q}$ is a measure of the stopping power of the medium
and has a similar importance for the characterization of the matter as does 
the shear viscosity for bulk transport.

\begin{figure}[tbp!]
\centering
\includegraphics[width=0.53\textwidth]{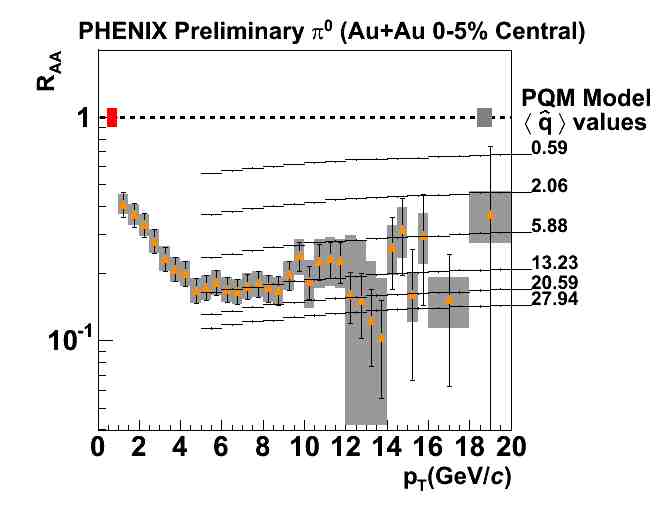}
\includegraphics[width=0.45\textwidth]{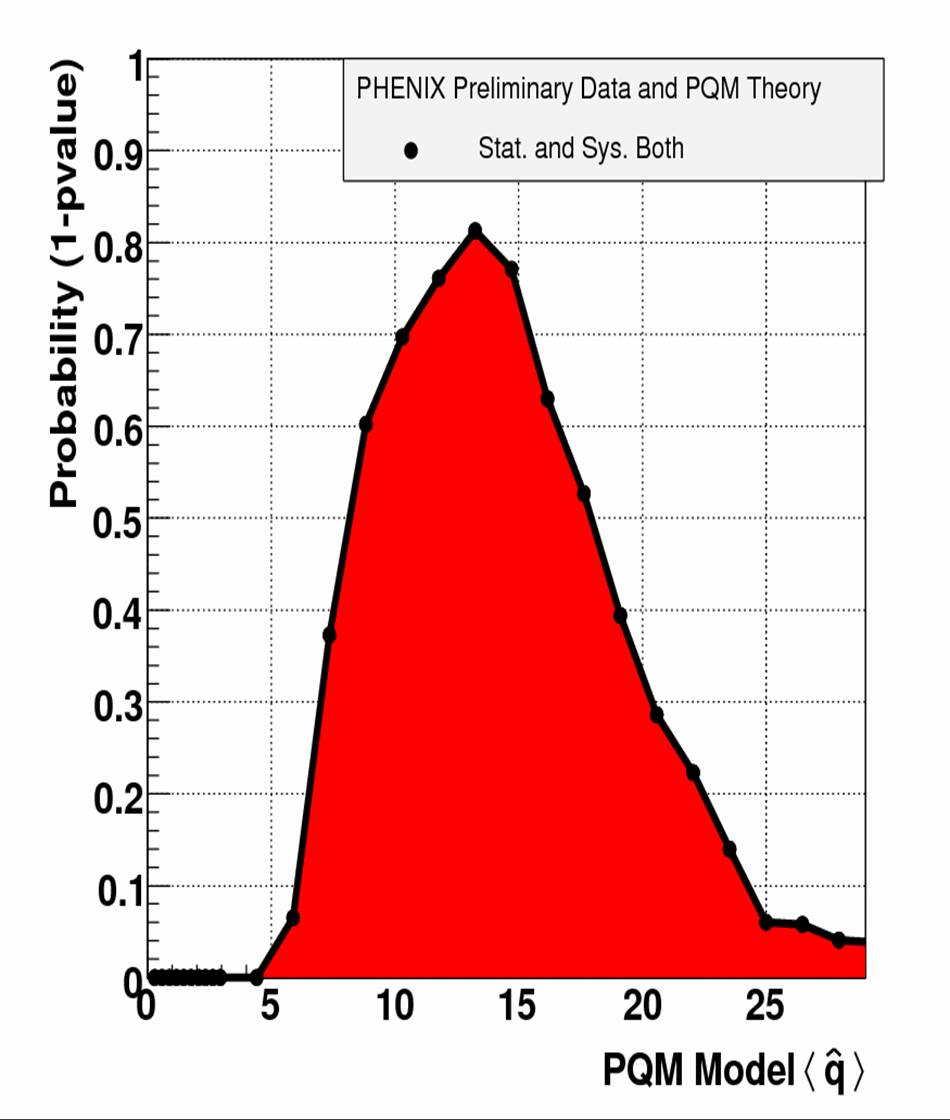}
\caption{Left: \RAA\ for \pizero\ compared to the PQM jet quenching model calculation 
for various values of model parameter \qhat. Right: $\chi^2$
probability distribution for \qhat. Figures from \cite{PhenixQM06}.
\label{fig:QhatPhenix}
}
\end{figure}

The relation between jet quenching observables (for sufficiently
energetic jets) and the parameter $\hat{q}$ is described within
perturbative QCD; the value of $\hat{q}$ itself is determined by
nonperturbative dynamics of the strongly interacting medium. An
important insight developed recently is that a nonperturbative and
gauge invariant definition can be given in terms of the expectation
value of a light-like Wilson loop. This definition has enabled new
calculational approaches to $\hat{q}$ (e.~g.\ in strongly coupled
QCD-like theories, see Section 2.3.3). Several groups have undertaken
detailed analyses of the RHIC data in terms of the parameter
$\hat{q}$, finding values more than 10 times larger than the stopping
power of normal nuclear matter. The results of these analyses still
differ considerably from each other, probably due to different approximations 
and oversimplifications made in the modeling of the collision geometry and
dynamics. A specific example of a fit to RHIC data on the suppression
of single hadrons and back-to-back hadron pairs is shown in
Fig.~\ref{fig:QhatPhenix}.

Detailed correlation measurements have shown that the yield of
high-$p_T$ particles correlated with the trigger particle but on the
opposite side is reduced by an additional factor of 4 or more, while
its energy is found to be carried away by enhanced production of soft
hadrons in the direction opposite to the trigger hadron
(Fig.~\ref{fig2.3}).  The average momentum of these soft hadrons
approaches that of the thermalized medium as the collisions become
more central and the fireball size increases. In addition, dramatic
modification in the angular and $p_T$ structure of the opposite side
jet point to a possible collective or hydrodynamic response of the
dense medium to the energy and momentum deposited by the quenched jet
(for example a Mach cone).

While not all of these features are understood quantitatively,
theoretical estimates of the initial gluon density present in the medium 
which are necessary to explain the observed high-$p_T$ hadron suppression 
are compatible with the values of the initial energy and entropy density 
required for the successful hydrodynamic description of the bulk of the matter. 
The important observation of an angular dependence of jet quenching relative 
to the reaction plane has opened the opportunity to use this process as a
tomographic probe for the properties of the dense medium created at RHIC.

\subsubsection{Novel Hadronization}

The kinetic freeze-out temperature $T_{\rm f}$ (determined by the
disappearance of elastic scattering) and the collective flow extracted
from the final hadron spectra depend on collision centrality. More
central collisions freeze out later, at lower temperature and with
larger radial flow than peripheral collisions, consistent with
theoretical ideas that describe kinetic freeze-out as a competition
between local scattering and global expansion rates. On the other
hand, the chemical decoupling temperature $T_{\rm ch}$ (defined by the
disappearance of abundance changing interactions) extracted from the
hadron yield ratios is found to be independent of collision centrality
and thus insensitive to the expansion rate. This observation, combined
with the value of $T_{\rm ch}$ near $T_c$, strongly suggests
that chemical freeze-out is not controlled by inelastic hadronic
rescattering processes, but by a phase change in which the hadrons are
born by a statistical process directly into a state which is
relatively dilute and expands so rapidly that most abundance-changing
hadronic interactions are ineffective.  Even hadrons with suppressed inelastic
interactions cross sections (for example the $\phi$ and $\Omega$)
follow the same freeze-out and flow patterns.

\begin{figure}[htbp]
\centering
\vskip -1cm
\includegraphics[width=0.9\textwidth]{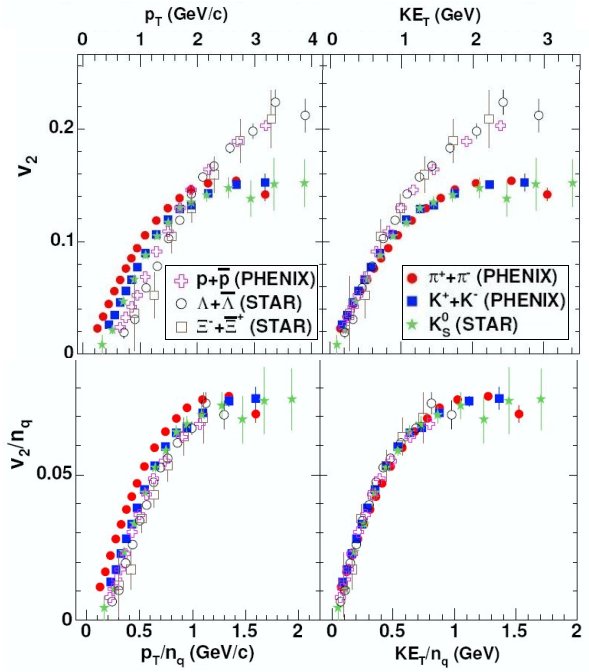}
\caption{Figure from \cite{V2Scale}.
Upper panels: The elliptic flow parameter $v_2$ plotted versus hadron
transverse momentum $p_T$ (left) or hadron transverse energy
$E_T=\sqrt{p_T^2+m^2}$ (right). At low transverse momentum/energy all
hadrons behave alike, indicating a common hydrodynamic origin of the
elliptic flow. At higher momentum/energy the data show a distinct
difference between mesons and baryons.  Lower panels: Elliptic flow
per valence quark $v_2/n_q$ versus transverse momentum per valence
quark $p_T/n_q$ (left) or transverse energy per valence quark
$E_T/n_q$.  The collapse of all data into a single curve in the lower
right panel indicates that the collective flow originates as a
hydrodynamical phenomenon at the valence quark level.
\label{fig:VtwoScaling}
}
\end{figure}

Further evidence for an active role of deconfined, thermalized and
collectively flowing quarks in hadron production comes from the
observed valence quark number scaling of hadron yields and elliptic
flow at intermediate $p_T$ (Fig.~\ref{fig:VtwoScaling}). While the
perfect liquid description gradually breaks down for $p_T \geq 1.5 - 2$
GeV/c, the broadening of the baryon spectra by the strong radial flow
remains visible at even larger transverse momenta.

A theoretical basis for the process of hadron formation at momenta in
the ``intermediate $p_T$'' range 2 GeV/c $<p_T<$ 5 GeV/c has been
developed. The model describes the formation of hadrons in this
momentum range as sudden recombination of collectively flowing valence
quarks to form mesons or baryons. This process imprints the
hydrodynamic flow characteristics of low-momentum quarks onto the
hadrons emitted with intermediate momenta. 

As a result, hydrodynamic
bulk particle production at low $p_T$ is separated from perturbative
hard particle production at high $p_T$ by a novel and unexpected
intermediate $p_T$ region where the parton recombination and
fragmentation mechanisms of hadron formation compete with each
other. The recombination model successfully describes the excess
emission of baryons at intermediate transverse momenta, the
characteristic difference between mesons and baryons in the momentum
dependence of the elliptic flow parameter, and the suppression of the
production of $p$-wave baryons.

\subsubsection{Saturated Gluon Density}

\begin{figure}[tbp!]
\centering
\includegraphics[width=0.9\textwidth]{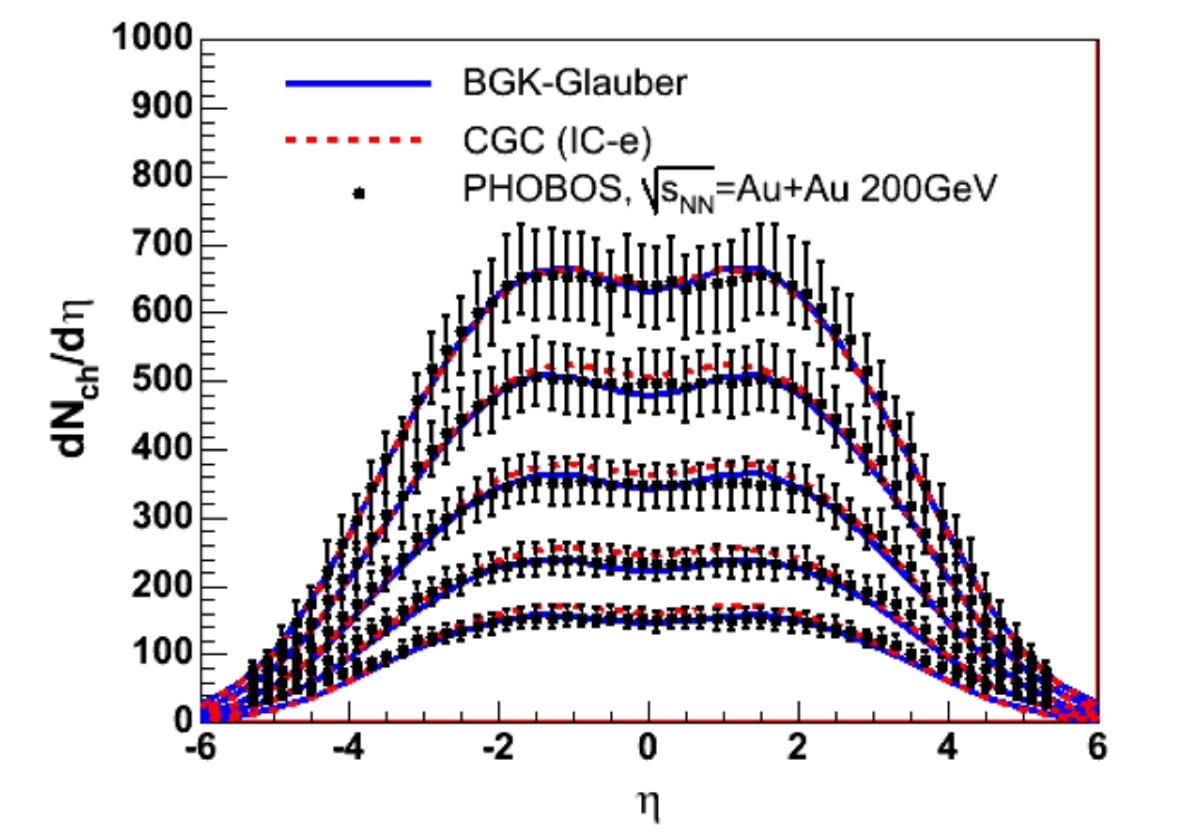}
\caption{dNch/d$\eta$ as function of pseudo-rapidity $\eta$ for variety of
 collision centralities \cite{PHOBOSdNdeta}, together with a fit using 
the Color Glass Condensate model in which the saturation of the density
of gluonic matter in the initial state leads to 
lower than expected particle multiplicity for 
central Au+Au collisions at RHIC energy.
\label{fig:dNdEta}}
\end{figure}

At high energies, the wave functions of hadrons and nuclei contain
many quarks and gluons -- this is because high energy (large $x$)
partons successively emit softer (smaller $x$) daughter partons in a
self--similar radiation cascade. Therefore, at small $x$ the density
of partons in the transverse plane becomes large. In this regime, the
softer gluons can recombine into harder ones, and this recombination
limits the growth of parton distributions, causing them to
saturate. The area density of partons defines a new dimensionful
parameter, the saturation momentum $Q_s$, which grows with the size of
the nucleus like $Q_s^2 \sim A^{1/3}$.  If the saturation momentum is
large compared to the confinement scale, asymptotic freedom dictates
that the coupling constant, and hence quantum effects, are small:
$\alpha_s(Q_s^2) \ll 1$. At $Q^2 \leq Q_s^2$ the dynamics of gluon
fields then becomes quasi-classical and highly non-linear. The
classical color fields in a highly energetic hadron or nucleus appear
frozen in time by Lorentz dilatation. This component of the wave
function is thus called the ``color glass condensate,'' and it is
predicted to become {\it universal}, i.e.\ the same for all hadrons
and nuclei, at very high energies.

This assumption can be probed in d+A collisions by concentrating on
kinematic regions sensitive to the small-$x$ gluon wave function of
the Au nucleus. Measurements performed at RHIC have, indeed, shown a
distribution of high transverse momentum particles at forward rapidity
whose dependence on rapidity, transverse momentum and centrality are
consistent with the Color Glass Condensate hypothesis. These
observations are the first indication that gluon saturation effects
play an important role in our understanding of nuclear structure at
small $x$ and of the pre-equilibrium stages of heavy ion collisions.
A second indication for the validity of the color glass condensate
picture is derived from the observed dependence of the particle
multiplicity in Au+Au collisions on centrality and beam energy
(Fig.~\ref{fig:dNdEta}), which can be understood as arising from such
a high-density state of gluonic matter present in the colliding Au
nuclei even at moderately small values of $x$.  This growing body of
evidence has led to the expectation that strong color fields will
determine the pre-equilibrium dynamics of heavy ion collisions at the
LHC. Many facets of this physics have unique manifestations also in
electron--nucleus scattering, which can be studied at a future
electron-ion collider (EIC).


\subsection{Theoretical Advances}

We have described many recent advances in theory in Section 2.2.
Here, we discuss four more theoretical developments that bear on the
interpretation of and context for RHIC data, as we build our
understanding of the phases of QCD matter.

\subsubsection{Lattice QCD at Finite Temperature and Density}

In recent years, significant progress has been made in studying the
phase diagram and bulk properties of QCD at finite temperature and
density. Previous lattice calculations were limited to zero net baryon
density, but several methods have recently been developed to study the
phase diagram, equation of state and various susceptibilities at
nonzero net baryon density. There is now solid evidence that the
transition from hadron gas to quark-gluon plasma at zero net baryon
density is a rapid crossover, not a true phase transition. However,
there exist general theoretical arguments and some indications from
lattice QCD that a critical end-point of a first-order transition line
exists at nonzero net baryon density.

\begin{figure}[tbp!]
\centering
\includegraphics[width=0.85\textwidth]{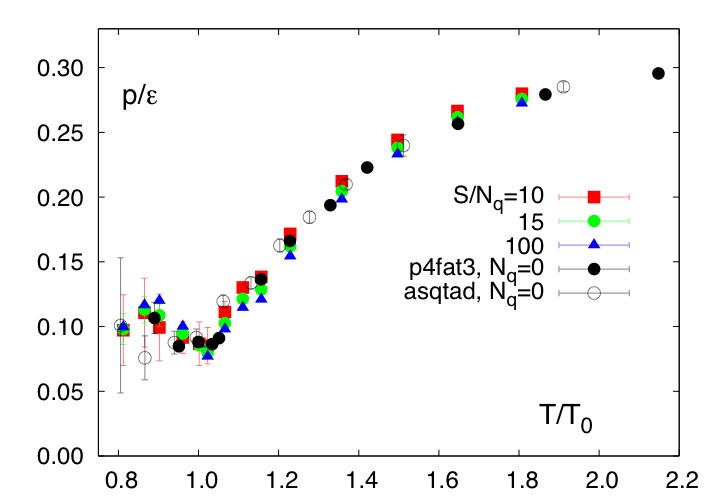}
\caption{Recent lattice results for the ratio $p/\varepsilon$ in 
unquenched QCD, which is an important input into the hydrodynamical
simulations of relativistic heavy ion collisions. The small value
$p/\varepsilon \ll 1/3$ near $T_c$ implies that the quark-gluon plasma
is characterized by a soft equation of state in the temperature range
relevant to RHIC \cite{Lattice1,Lattice2}.
\label{fig:soft_eos}}
\end{figure}

For the first time the transition temperature in QCD has been
calculated with controlled continuum and chiral extrapolations using
the improved staggered fermion action.  One recent calculation of the
transition temperature gives $T_c=192(7)(4)$MeV which is larger than
other published values of $T_c$ and the chemical freezout temperature. 
If confirmed, this result would have important implications for the phenomenology 
of heavy-ion collisions.  The calculations of the equation of state can now be
performed with quark masses near their physical values, and definitive
results for thermodynamic quantities with dynamical quarks in the
continuum limit appear to be in reach.

Figure~\ref{fig:soft_eos} shows results based on recent calculations
of the QCD equation of state with dynamical fermions described by an
improved lattice action. The figure shows the ratio of the pressure
$p$ to the energy density $\varepsilon$ as a function of the
temperature. This quantity, which would have the value 1/3 for a
perturbative gas of massless quarks and gluons or for a strongly
coupled liquid that is conformal (i.e. scale-invariant), is closely
related to the speed of sound $c_s$ in the plasma ($c_s^2 =
\partial p/\partial\varepsilon$) and is thus relevant to 
hydrodynamical calculations of the expansion of the matter formed in
heavy ion collisions.

\subsubsection{Lattice Spectral Functions}
\label{sect:LatticeSpectFn}

For a long time lattice QCD has been used only to calculate static
properties of the quark-gluon plasma, such as the transition temperature,
equation of state and screening lengths. In recent years significant
progress has been made in calculating the temporal meson correlators
and spectral functions using the {\em Maximum Entropy
Method}. Charmonium spectral functions have been calculated by several
groups indicating that the ground state charmonium ($J/\psi$) can
survive up to temperatures at least as high as $1.6\ T_c$.  These
findings differ from early estimates based on the perturbative
color screening scenario, which predicted that the charmonium ground
state would dissolve soon above $T_c$. Charmonium correlators
calculated on the lattice thus become an essential input for
phenomenological models aimed at understanding charmonium production
at RHIC and for any theoretical approach addressing color screening in
the quark-gluon plasma.

\subsubsection{Strong coupling results from AdS/CFT correspondence}

There is now a significant body of experimental evidence for the
discovery that the quark-gluon plasma produced in RHIC collisions is a
strongly coupled liquid with low viscosity, not a near-ideal gas.
Lattice QCD is the proper tool for understanding the equilibrium
thermodynamics of such a strongly coupled quark-gluon plasma, but the
discovery poses a challenge to the theoretical analysis of its
transport properties. The theoretical tools of choice for the
understanding of observable phenomena, such as the strong jet
quenching, which involve dynamics rather than just thermodynamics, have
long been built upon perturbative QCD and are thus based upon the
premise that interactions are essentially weak.  This assumption does 
not apply to the matter produced at RHIC. The search for new tools to 
study the transport properties of matter described by strongly coupled, 
relativistic gauge theories has thus become an urgent necessity.

Recently, theorists have calculated the shear viscosity/entropy ratio
$\eta/s$, the jet quenching parameter $\hat q$, the drag coefficient
describing the energy loss of a heavy quark, the photon emission rate,
and the velocity dependence of the color screening length for the
strongly coupled plasmas of many gauge theories that differ in
detail from QCD. The calculations are made possible by the fact that
large classes of strongly coupled, thermal gauge theories are
equivalent to string theories in curved 5-dimensional space-times
containing black holes.  This ``AdS/CFT correspondence'' was
discovered by string theorists hoping to use gauge theories to learn
about string theory. Nuclear theorists are now putting it to
profitable use in the opposite direction.  These calculations yield
new insights: for example, there may be a fundamental lower bound on
the ratio of shear viscosity $\eta$ to entropy density $s$; for
example, $\hat q$ scales with the square root of the number of degrees
of freedom; for example, heavy quark energy loss may occur via drag
rather than via the gluon radiation which dominates in the
high-jet-energy limit and which is described by $\hat q$.  In several
instances ($\eta/s$ and $\hat q/T^3$, for example) the AdS/CFT results
obtained at strong coupling yield results which are in
semi-quantitative agreement with those inferred from RHIC data, even
though they are not calculations done in QCD.

\subsubsection{Cold Dense Quark Matter}

Theoretical advances have shown that QCD provides rigorous analytical
answers, leaving no unresolved gaps in our understanding even at a 
nonperturbative level, to the question: 
``What are the properties of matter squeezed to arbitrarily high density?''  
It has long been known that cold dense
quark matter, as may occur at the center of neutron stars, must be a
color superconductor.  Recent theoretical effort has made this subject
both richer and more quantitative.  An analytic, ab-initio calculation
of the pairing gap and critical temperature at very high densities has
now been done, and the properties of quark matter at these densities
have been determined. The material is a color superconductor but
admits a massless ``photon'' and behaves as a transparent insulator;
it is a superfluid with spontaneously broken chiral symmetry.  At
densities that are lower but still above that of deconfinement, color
superconducting quark matter may in a particular sense be crystalline,
with a rigidity several orders of magnitude greater than that of a
conventional neutron star crust.

\section{Phases of QCD: Future Prospects}
\label{HIFuture}

We are poised at the beginning of a new era in the quantitative
experimental exploration of thermal QCD. This is made possible by
dramatic detector and accelerator advances at RHIC and the opening of
a new energy frontier at the LHC, which extends the experimental
exploration of the phase diagram to yet higher temperatures. RHIC (and
eventually the FAIR facility) will also explore the new region of
finite baryon density, where lattice QCD calculations predict a
critical point that is potentially accessible to RHIC.

This chapter presents the plans of the Heavy Ion community to address
the challenges and opportunities. We first discuss the upgrades of the
RHIC accelerator and detectors and the heavy ion capabilities of the
LHC detectors, followed by discussion of important aspects of the
physics scope of these upgraded and new facilities.

\subsection{Facilities}

\subsubsection{RHIC}

The initial suite of RHIC detectors comprised two large, general
purpose experiments (PHENIX and STAR) and two small specialized
experiments (BRAHMS and PHOBOS). BRAHMS and PHOBOS have completed
their physics programs. Substantial upgrades to the PHENIX and STAR
detectors are now in progress, at a total cost of about \$30M. These
upgrades will enable the detectors to address the key questions
enumerated in this document, through extended particle identification
capabilities (including heavy flavor mesons and baryons) and kinematic
coverage, as well as improved triggering and data recording
capabilities. Many new measurements require large data samples, to
have sensitivity to processes that occur at the level of once per
hundred million Au+Au reactions. The RHIC accelerator complex is also
being upgraded, in response to the physics needs for high luminosity
and a broader range of available species and energies. This upgrade
program is detailed in the ``Mid-Term Strategic Plan for RHIC''
\cite{RHICMidtermPlan}. 

\ \\
\noindent
The major detector upgrades for {\bf PHENIX} are:

\begin{itemize}

\item Hadron Blind Detector: Ring Imaging Cerenkov detector for 
high signal/background measurements of low mass electron pairs, to
study thermal radiation and medium-induced modification of mesons,
which may be sensitive to chiral symmetry restoration;

\item Central and forward silicon trackers: 
high precision detectors for resolved-vertex studies of heavy flavor
production and electron-pair production over broad acceptance;

\item Muon trigger upgrade: enhanced capability to trigger on 
$W^\pm\rightarrow\mu^\pm$, 
to measure the sea-quark contribution to nucleon spin;

\item Nose-Cone Calorimeter: a forward tungsten-silicon calorimeter measuring 
photons and electrons, to study heavy quark spectroscopy and forward
jet production at low $x$.

\item Data acquisition upgrade to accomodate high data volume 
from new detectors

\end{itemize}

\ \\
\noindent
The major detector upgrades for {\bf STAR} are:

\begin{itemize}

\item Forward Meson Spectrometer: large acceptance forward 
lead-scintillator calorimeter to measure forward meson, photon, and
heavy quark production at low $x$;

\item Time of Flight: highly segmented, MRPC-based detector 
covering the central STAR acceptance, to identify hadrons and
electrons over broad kinematic interval; substantially improved
measurements of event-by-event fluctuations and heavy flavor and
vector meson production;

\item Heavy Flavor Tracker: high precision silicon 
detectors for resolved-vertex studies of heavy flavor production and
electron-pair production over broad acceptance;

\item Forward Tracker: GEM-based detector in same acceptance as STAR 
Endcap EM Calorimeter, to measure the sea-quark contribution to
nucleon spin via $W^\pm\rightarrow{e}^\pm$;

\item High-speed Data Acquisition: increase of STAR event readout speed 
by a factor 10, to 1 kHz, to enable recording of very large datasets
needed for high precision event-by-event and heavy flavor studies.

\end{itemize}

\begin{figure}[tbp!]
\centering
\includegraphics[height=0.45\textheight]{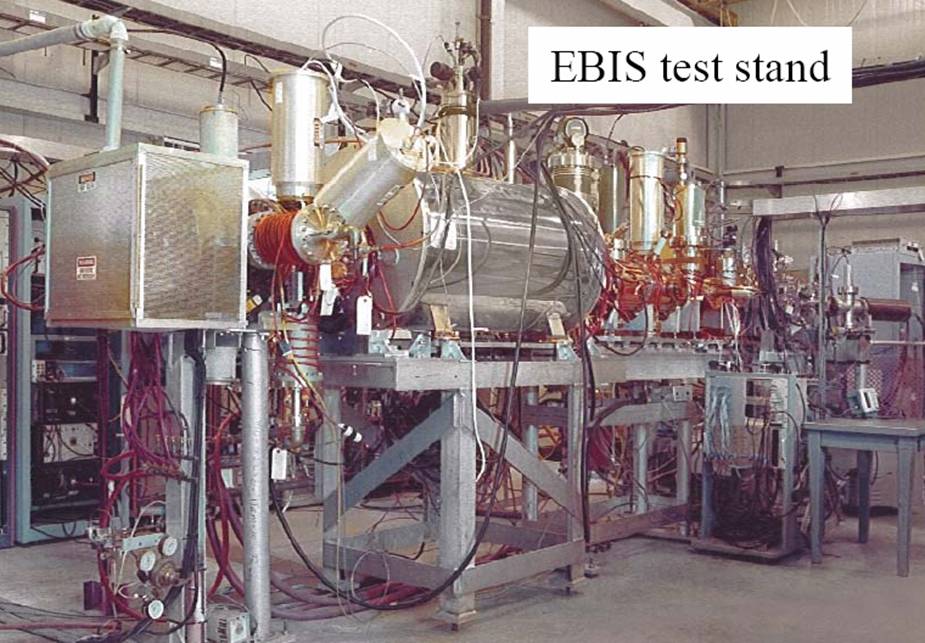}
\vskip 1.25cm
\includegraphics[width=0.95\textwidth]{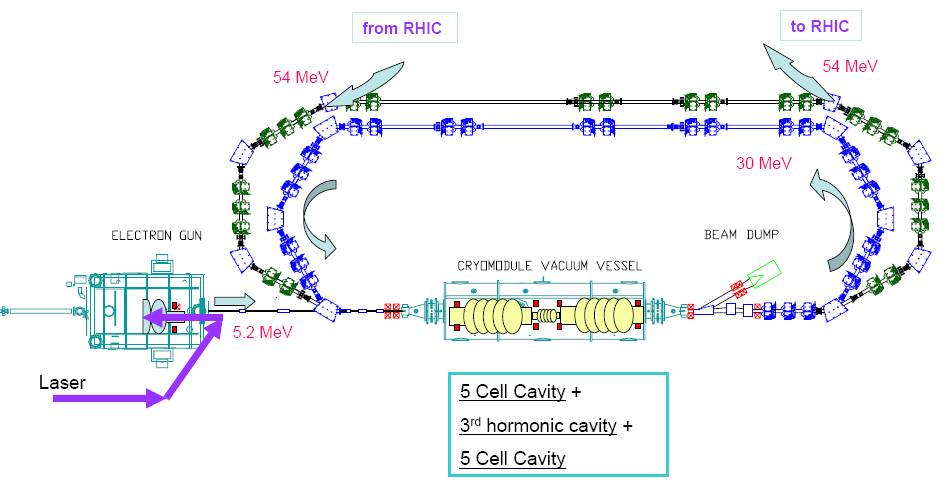}
\vskip 0.75cm
\caption{Upper: EBIS test stand. Lower: Schematic layout of electron cooling. 
Location in RHIC ring is shown in Fig. \ref{fig:EIC}.
\label{fig:RHICUpgrades}}
\end{figure}

The short term upgrade to the RHIC accelerator facility is the
Electron Beam Ion Source (EBIS) (Fig. \ref{fig:RHICUpgrades}, upper
panel). EBIS is currently under construction and will be commissioned
and operational in 2010. It will replace the 35-year-old Tandem Van de
Graaffs as the RHIC ion source, providing more reliable and
cost-effective operation. EBIS makes new species available in RHIC,
notably polarized Helium-3 and Uranium. Measurements of U+U collisions
are of particular interest, because the large ground-state quadrupole
deformation can be exploited to generate initial energy density about
30\% higher than that achievable in Au+Au collisions at RHIC. This
lever arm may provide significant systematic checks of hydrodynamic
flow and jet quenching.

As discussed in Sect.~\ref{QCDCriticalPoint}, one of the key open
questions for the field is the existence and location of a critical
point on the QCD phase diagram. RHIC will explore this question,
varying the baryo-chemical potential $\mu_B$ by lowering the collision
energy as far down as $\sqrt{s_{NN}}=5$ GeV. Effective use of such an
energy scan, for instance to measure the event-by-event fluctuations
that characterize the vicinity of the critical point at multiple
energies within a single running period, requires the completion of
detector upgrades underway, in particular the STAR Time of Flight
upgrade. Discoveries from such a scan could be studied in greater
detail with substantial improvement to the luminosity at low
energy, which can be achieved at moderate cost through electron
cooling in the AGS. This would increase the luminosity below
$\sqrt{s_{NN}}\sim20$ GeV by a factor up to 30.

In the longer term, the required high luminosity at RHIC II will be
achieved by electron cooling of the full energy beams
(Fig. \ref{fig:RHICUpgrades}, lower panel), the first such
implementation in a high energy collider. Cooling will increase the
heavy ion luminosity by a factor 10 at high energy and make RHIC the
first collider in which luminosity is limited by the interactions
themselves. Cooling at injection energy will increase polarized proton
luminosity by a factor 2-3. Proof of principle of electron cooling at
RHIC has been established through detailed simulations benchmarked at
the existing high energy electron cooler at Fermilab. Major components
of the RHIC electron cooler will be tested in a scaled test facility
that is currently under construction. Commissioning of the full system
could technically be completed by 2012.

RHIC computing capabilities must accomodate the large increase in data
volume and complexity resulting from the increased RHIC II luminosity
and upgrades to detectors. Detailed estimates indicate that the
expected decrease in the cost of computing capacity with time offsets
the increase in demand. Consequently, no significant increase in
funding level for computing is required for the on-site RHIC Computing
Facility and off-site satelite facilities to keep pace, in order to
analyze the data in a timely fashion.

\subsubsection{LHC}

The LHC physics program includes four weeks of heavy ion physics
running per year. The primary collision system is Pb+Pb at 5.5 TeV per
nucleon pair, which is a factor 30 greater collision energy than at
RHIC. Other systems under consideration are p+Pb (achievable at LHC
despite the two-in-one magnet design) and 5.5 TeV p+p, to provide
accurate reference data for heavy ion collisions measurements. The LHC
is currently expected to begin commissioning with proton collisions in
late 2007, with heavy ion beams commisioned in late 2008 and the first
significant heavy ion running in 2009.

LHC heavy ion collisions are expected to generate matter with much
higher initial energy density than RHIC collisions, with a long-lived
fireball in the deconfined phase. The enormous collision energy
results in large rates for a wide variety of hard probes over a very
broad kinematic range (for instance, jets exceeding 300 GeV in
transverse energy), which will complement and extend the successful
hard probes measurements at RHIC and will in addition enable qualitatively new
measurements.

Three LHC experiments - ALICE, ATLAS and CMS - will participate in
heavy ion running, with extensive capabilities to measure the full
spectrum of heavy ion observables. The ability to study similar probes
of hot QCD matter generated from the vastly different initial
conditions at RHIC and the LHC promises a rich physics program for the
two facilities in the coming years.

We outline here the heavy ion physics capabilities of the LHC
detectors:

\begin{itemize}

\item {\bf ALICE} is the dedicated heavy ion experiment at the LHC. 
ALICE contains the main elements of both STAR and PHENIX. Its central
detector, with acceptance $|\eta|<0.9$, has a large Time Projection
Chamber in a moderate solenoidal field (0.5 T), augmented by silicon
tracking and highly segmented electron and hadron particle ID
detectors. A muon arm in the forward direction is based on a
large-aperture dipole. The US hardware contribution to ALICE is a
large Electromagnetic Calorimeter covering one third of the central
acceptance, which enables jet quenching measurements in ALICE over a
broad kinematic range.

\item {\bf ATLAS} is a large acceptance, multi-purpose detector, 
with silicon and TRD-based tracking and highly segmented
electromagnetic and hadronic calorimeters within a 2 T solenoid
magnet. Muons are detected in large air-core toroids surrounding the
central detector. US groups provide the physics leadership of the
overall ATLAS Heavy Ion effort, while the US heavy ion hardware
contribution is a modest-scale project to provide the Zero-Degree
Calorimeters.

\item {\bf CMS} is a large acceptance, multi-purpose detector, 
with silicon-based tracking and and highly granular electromagnetic
and hadronic calorimeters within a 4 T solenoid magnet. Muon detectors
are embedded in the flux return iron yoke of the magnet. The very
forward direction is covered by CASTOR and the Zero-Degree
Calorimeters. US insitutions provide physics leadership of the overall
CMS heavy-ion physics program, trigger preparation, and ZDC
construction.

\end{itemize}

While ALICE is the only LHC detector that was designed from the outset
for high performance tracking the high multiplicity environment
expected in heavy ion collisions, subsequent studies of CMS and ATLAS
have shown that they can also track robustly in such an
environment. All experiments have good capabilities for heavy ion jet
quenching and photon and quarkonium production measurements. Dimuon
mass resolution is expected to be sufficient to separate the various
quarkonium states. Extensive forward coverage will enable jet and
photon measurements at moderate $Q^2$ down to $x\sim10^{-6}$ (ATLAS)
or even $x\sim 10^{-7}$ (CMS), which is of particular interest in p+Pb
collisions.

\subsection{The QCD Critical Point}
\label{QCDCriticalPoint}

Measurements using the RHIC detectors and complementary lattice
calculations that each extend into the regime of nonzero baryon
density can revolutionize our quantitative understanding of the QCD
phase diagram by discovering the QCD critical point.

What is the nature of the transition between Quark-Gluon Plasma and
ordinary hadronic matter?  Lattice calculations show that in a
matter-antimatter symmetric environment, this transition occurs
smoothly, with many thermodynamic properties of QCD matter changing
dramatically within a narrow range of temperatures, but with all these
changes occurring continuously. Collisions at the highest RHIC
energies produce matter that is close to matter-antimatter symmetric,
as did the big bang. RHIC data and cosmological observations are
consistent with the prediction that the transition undergone by
quark-gluon plasma as it cools occurs continuously.

\begin{figure}[tbp!]
\includegraphics[width=0.55\textwidth]{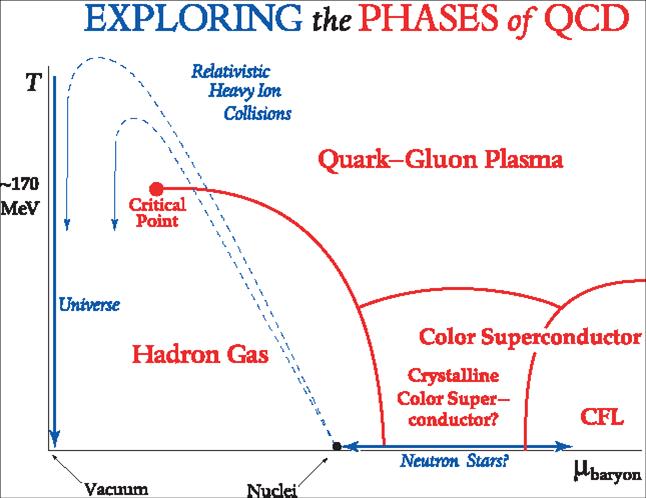}
\includegraphics[width=0.45\textwidth]{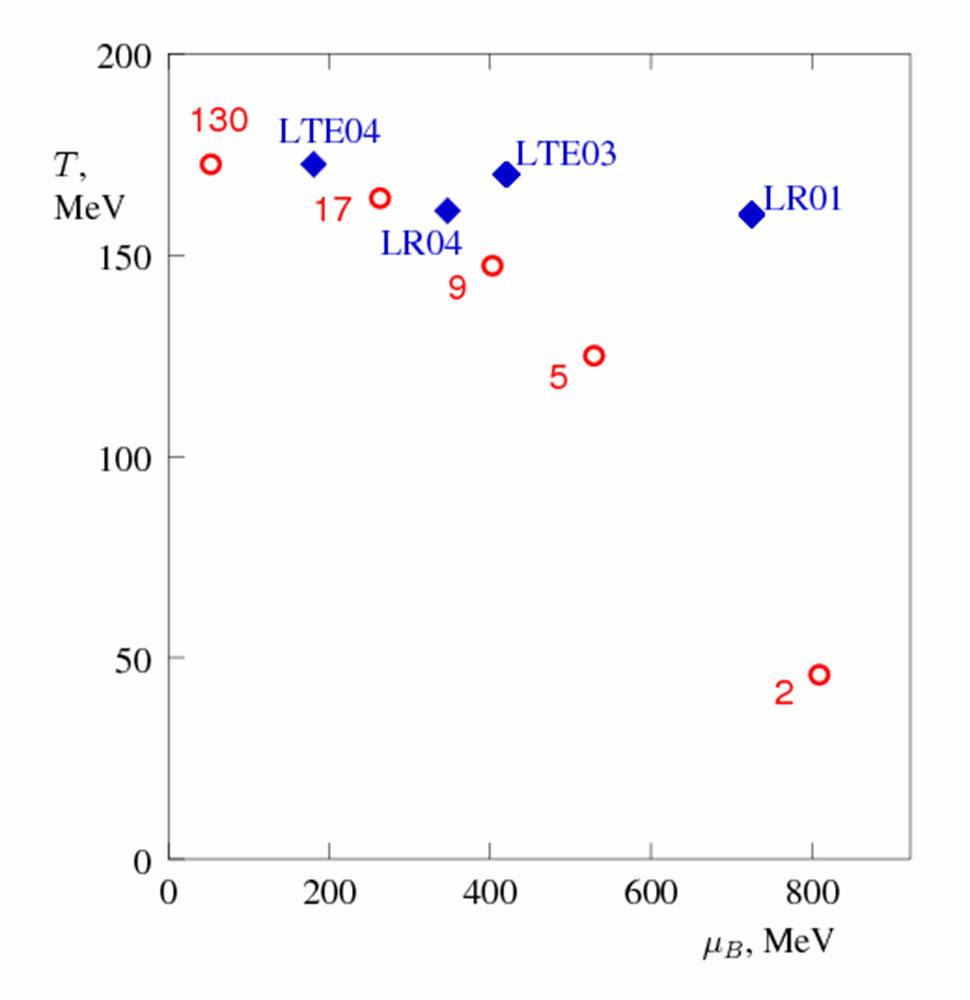}
\caption{{\bf Left panel:} a sketch of the QCD phase diagram as a function of temperature
$T$ and baryon chemical potential $\mu_B$.  The early universe cooled
slowly down the vertical axis --- it was filled with quark-gluon
plasma for the first microseconds after the big bang.  Heavy ion
collisions reproduce matter last seen in nature at this early
cosmological epoch.  The transition between quark-gluon plasma and
ordinary hadronic matter is a crossover at small $\mu_B$, and is
thought to become first order for $\mu_B$ greater than that of a
critical point in the phase diagram.  Cold dense quark matter, as may
occur within neutron stars, is in one of several possible color
superconducting phases. {\bf Right panel:} searching for the QCD
critical point \cite{Stephanov}. 
The blue diamonds mark the location of the critical
point found in four pioneering lattice QCD calculations (done in the
years indicated, using the ``Lattice Reweighting'' or ``Lattice Taylor
Expansion'' methods). Each was done at a single lattice spacing;
extrapolation to the continuum limit is a current challenge.  The red
circles, labeled by $\sqrt{s}$, indicate the location in the phase
diagram where heavy ion collisions with various collision energies
freeze out.  By scanning $\sqrt{s}$ over a range extending down to 5
GeV, and by virtue of its favorable collider geometry and detectors,
RHIC can look for the entire suite of event-by-event fluctuations
expected to characterize collisions which freezeout after passing near
the critical point if the critical point has
$\mu_B<\sim500$~MeV.
\label{fig:PhaseDiagram}
}
\end{figure}

In contrast, upon squeezing nuclear matter to higher and higher
densities without heating it up --- a feat accomplished in nature
within the cores of neutron stars --- we expect one or more first
order phase transitions (at which thermodynamic properties change
discontinuously as the pressure is increased) between various phases
of nuclear matter and color superconducting quark matter
(Fig.~\ref{fig:PhaseDiagram}, left panel). Furthermore, the phase transition
between cold dense color superconducting quark matter and hot
quark-gluon plasma must, on very general grounds, be a first order
transition.

Many studies which seek to put these facts and expectations together
into a map of the QCD phase diagram predict that the continuous
crossover being explored in heavy ion collisions at the highest RHIC
energies will become discontinuous if the excess of matter over
antimatter, typically parametrized by a chemical potential $\mu_B$ for
baryon number, can be increased above some critical value.  The
critical point where the transition changes its character is a
fundamental landmark on the phase diagram of QCD. Within the last five
years, new methods in lattice QCD have opened the door to an {\it ab
initio} theoretical determination of its location
(Fig.~\ref{fig:PhaseDiagram}, right panel). If it lies at a $\mu_B$ which
is not more than 500 MeV, as several of the pioneering lattice
calculations indicate, then there is every expectation that as these
calculations are pushed to finer lattice spacings over the coming few
years the QCD prediction for the location of the critical point will
become as solid as that for $T_c$, the temperature of the
matter-antimatter-symmetric crossover, is today.

At present we have only a tentative sketch of the QCD phase diagram
based on what we know at $\mu_B=0$, together with models, inferences,
and the pioneering lattice QCD calculations. The discovery of
experimental evidence for the existence, and hence the location in
$(\mu_B,T)$, of either the QCD critical point itself or the first
order phase transition that lies beyond it at higher $\mu_B$, would
transform this sketch into a solid, quantitative map in which we have
full confidence. Increasing $\mu_B$ (i.e. increasing the excess of
matter over antimatter) in a heavy ion collision is accomplished by
reducing the collision energy $\sqrt{s}$.  By doing heavy ion
collisions at a sequence of energies with $\sqrt{s}$ between 5 and 50
GeV, RHIC will be able to explore the character of the QCD transition
with $\mu_B$ between 30 and about 550 MeV, meaning that it can find
the QCD critical point if it lies within this broad regime.

Heavy ion collisions that cool in the vicinity of the QCD critical
point are expected to be characterized by enhanced event-by-event
fluctuations of all observables that depend either on the
matter-antimatter asymmetry or on the degree of chiral symmetry
breaking. Both these quantities fluctuate with large amplitudes and
over long length scales only near the critical point, and many
properties of these fluctuations can be calculated from first
principles.  Examples of observables which have been studied include
the event-by-event fluctuations of the number of protons minus
antiprotons, of the mean transverse momentum $p_T$ of all the soft
pions in an event, and of the kaon-to-pion and proton-to-pion ratios.

Earlier experiments at the SPS have found an intriguing and as yet
unexplained enhancement in the fluctuations of the kaon-to-pion ratio
at $\sqrt{s}=6-8$~MeV, corresponding to $\mu_B=400-500$~MeV.  The
proton-to-pion fluctuations are not enhanced, and the $p_T$
fluctuations were not measured.  RHIC will be able to make comparative
measurements over a range of energies in collider geometry, and with
the same detectors, offering a considerable advantage because most
systematic effects will remain constant, in contrast to the situation
for fixed target measurements. Recent studies indicate that, in a
single running period, RHIC could significantly improve both the
statistical and systematic errors on those observables where there are
existing data at SPS energies, while at the same time making the whole
suite of relevant event-by-event fluctuation measurements over the
entire relevant energy range for the first time. The feasibility of
this program depends crucially on the completion of detector upgrades
currently underway and depends on the capability of RHIC to provide
adequate luminosity at low $\sqrt{s}$.  There was considerable
discussion of these feasibility issues at a recent workshop held at
BNL \cite{BNLCritPoint}.  It was concluded that there are no apparent
barriers for operation of RHIC at 5-50 GeV, allowing RHIC to access
the entire range 30 MeV$<\mu_B<$550 MeV.  The possibility of
implementing electron cooling to increase the luminosity at low
energies is being pursued. This would make it possible to study any
newly discovered features of the QCD phase diagram with greater
precision, for example by permitting the high-statistics runs needed
to see dileptons or by facilitating varying nuclear size $A$ along
with $\sqrt{s}$.

When the FAIR facility, with its CBM detector, comes on line in
Germany in 2015, it will study matter with $\sim400 < \mu_B <$650 MeV. If RHIC
discovers the QCD critical point, experiments at FAIR will be
well-positioned to study the spatially inhomogeneous final state of
heavy ion collisions which cool through a first order phase
transition.  If RHIC discovers that the QCD critical point lies at
$\mu_B>400$ MeV, the FAIR facility will seek to confirm this discovery
directly.

Locating the critical point where the transition changes its character
is of fundamental importance for understanding QCD.  This provides
RHIC with significant new discovery potential as it explores the
poorly charted reaches of our current map of the QCD phase diagram.
As is the case in the complementary effort to use RHIC to gain new and
more quantitative understanding of the properties of quark-gluon
plasma, ramping up corresponding theoretical efforts are crucial to
the success of the experimental program.  The experimental search for
the QCD critical point will require the phenomenological studies of
the properties and experimental signatures of matter near the critical
point to be taken to a new level, ultimately in an interplay with data
as it comes in.  Furthermore, in order for an experimental discovery
to have maximum impact, the recent advances in lattice QCD that have
opened this regime to {\it ab initio} calculations must be pursued and
capitalized upon. Lattice QCD calculations and RHIC experiments are in
a race to locate the QCD critical point. The resulting scientific
accomplishment will have the greatest impact only if both parties win.

\subsection{Hard Probes at RHIC and LHC}

Hard (high \Qsq) probes, in particular energetic jets, heavy quarks,
and quarkonia, have provided key insights into the QCD matter
generated in high energy nuclear collisions
(Sect.~\ref{JetQuenchingStatus}). The great utility of hard probes for
quantitative measurement of the transport properties of dense matter
is due to several factors:

\begin{itemize}

\item hard probe production rates in nuclear collisions can be established 
using perturbative calculations and measurements in p+p and p/d+A
collisions;

\item interactions of hard probes in dense matter are seen experimentally to be very strong; 

\item these interactions are theoretically calculable within perturbatively-based frameworks.

\end {itemize}

The most notable success thus far in the area of hard probes at RHIC
is jet quenching. Measurements of quarkonium suppression are now
coming to maturity and promise as great an impact on our understanding
of QCD matter. Here we discuss prospects for both classes of
measurement in the coming years at RHIC and the LHC, and the
developments necessary to realize them.

\subsubsection{Jet Quenching}
\label{FutureJetQuenching}

Measurements of high \pT\ hadron production and correlations give
direct evidence that extraordinarily large gluon densities (and
correspondingly large energy densities) are generated in head-on
collisions of heavy nuclei (Sect. \ref{JetQuenchingStatus}). However,
quantitative interpretation of these data in terms of gluon densities
and transport properties of the medium requires comparison to detailed
phenomenological models. The systematic uncertainties of the extracted
quantities therefore depend not only on the precision of the data, but
also on the validation of the theoretical models and the full
exploration of their parameter space. As the following discussion
illustrates, jet quenching provides an example of striking
experimental discoveries whose importamce and implications can be
fully realized only by a combination of more discriminating,
differential measurements and significant progress in theoretical
understanding of the underlying processes.

\ \\
{\bf Measurement of transport properties:} Significant recent progress
has been made in the quantitative comparison of jet quenching data
with detailed theoretical calculations. Models incorporating radiative
energy loss reproduce accurately the systematic behavior of hadron
production and correlations as a function of hadron momentum and
system size. High \pT\ pion suppression constrains the transport
coefficient \qhat\ in such models to within a factor two
(Fig.~\ref{fig:QhatPhenix}), while the additional consideration of
back-to-back leading hadron suppression from a pair of recoiling jets
can reduce the systematic uncertainty further. However, due to
the large partonic energy loss, the core of the fireball is largely
opaque to moderate energy jets. Precise determination of medium
properties in this framework requires systematic checks over a much
broader dynamic range, which are possible only for dihadron
correlation measurements at substantially higher momentum. Such
measurements require RHIC II luminosities and the higher collision
energy at the LHC.

\begin{figure}[tbp!]
\centering
\vskip -0.5cm
\includegraphics[width=0.95\textwidth, height=10 cm]{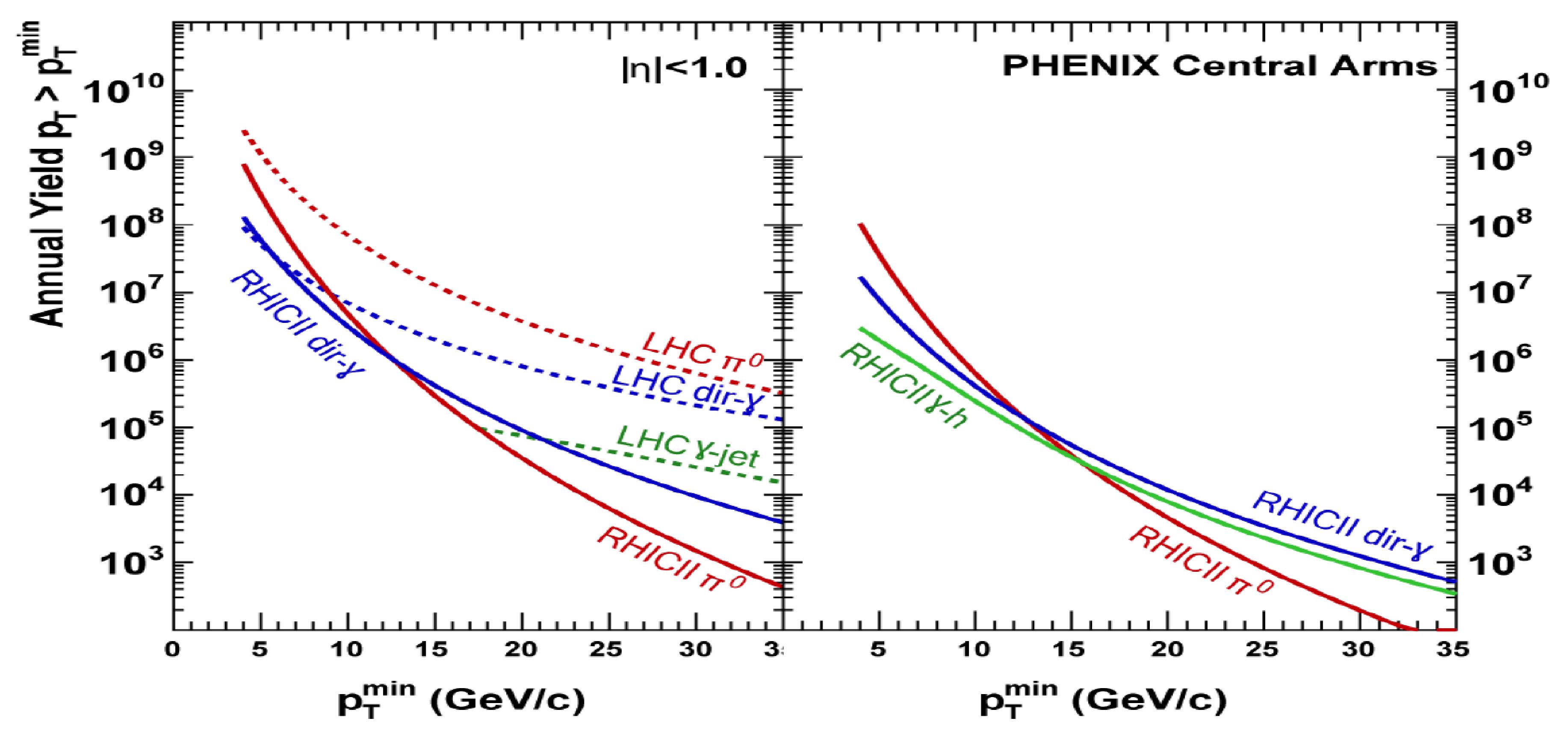}
\caption{Annual yields at RHIC II and LHC for \pizero, direct $\gamma$, 
and $\gamma$+jet above a \pT\ cut, for STAR (left) and PHENIX (right)
acceptances. ATLAS and CMS have larger acceptance than left panel,
ALICE has slightly smaller acceptance. Yields from RHIC I, prior to
the luminosity upgrade, are a factor $\sim10$ smaller than RHIC II yields.
\label{fig:GammaJet}
}
\end{figure}

These results are striking, and represent a major success of the RHIC
program. However, the primacy of radiative energy loss is challenged
by measurements of non-photonic electrons, which arise from the
semi-leptonic decay of heavy flavor mesons (charm and bottom). The
magnitude of radiative energy loss is expected on general grounds to
be reduced for massive quarks, with a large difference predicted
between charm and bottom quarks in the kinematic range currently
accessible for non-photonic electron measurements at RHIC ($\pT<10$
GeV/c). Surprisingly, RHIC measurements show that non-photonic
electron production in central nuclear collisions is suppressed at the
same level as light hadron production (a factor $\sim5$), leading to
an apparent inconsistency for calculations in which only radiative
energy loss is considered.

Resolution of the heavy flavor suppression puzzle at RHIC is crucial
to establish a fully self-consistent picture of jet quenching. One
proposed solution is the introduction of additional energy loss
mechanisms, in particular elastic channels. However, the magnitude of
the actual discrepancy depends on the relative contribution of charm
and bottom mesons to the non-photonic electron yield, which at present
is not well known. Perturbative QCD calculations unfortunately do not
provide meaningful constraints on these contributions.  The essential
missing ingredient is the measurement of charm and bottom suppression
separately, which requires the vertex detector upgrades to PHENIX and
STAR that are currently in progress.

\ \\
{\bf $\gamma$+jet and $Z$+jet measurements:} Due to large partonic
energy loss, the core of the fireball is effectively opaque to jets,
and the hadronic measurements described above are dominated by jets
generated at its periphery. This geometric bias limits sensitivity to
the hottest and densest matter at the core of the fireball, and
transport properties deduced from such measurements therefore have
significant model dependence. The QCD Compton process, where a jet
recoils from a hard direct photon (or $Z$-boson at the LHC), does not
suffer from this geometric bias since the trigger photon/$Z$ does not
carry color charge and therefore does not lose energy in the
medium. This process probes the full volume of the fireball, with the
direct $\gamma$/$Z$ providing an accurate measurement of the energy of
the recoiling jet. Though challenging in terms of rate and
signal/background, these measurements will provide the most precise,
model-independent measurements of jet quenching available. Both PHENIX
and STAR have reported initial steps in this direction, but current
data are very limited statistically. High quality measurements require
a luminosity upgrade. Figure \ref{fig:GammaJet} shows the annual yield
of direct photons, \pizero{s} and $\gamma$+jet/hadron coincidences in
heavy ion collisions at RHIC II and the LHC. The kinematic reach of
the RHIC detectors following the RHIC II luminosity upgrade is
significant for $\gamma$+hadron coincidences, extending well beyond 30
GeV/c. At the LHC a similar range is expected for statistically
significant $Z$+jet measurements, while the $\gamma$+jet measurement
will extend farther.

\ \\
{\bf Novel jet quenching phenomena:} As discussed in
Section~\ref{JetQuenchingStatus}, the detailed study of jet structure
in nuclear collisions continues to reveal surprising new
phenomena. Low momentum particles recoiling from a trigger hadron are
distributed in a broad cone, perhaps indicating the generation of
shock waves or Cerenkov radiation in the medium. Intra-jet
correlations, measured using hadron pairs at small angular separation,
are elongated in the beam direction, perhaps due to coupling of
medium-induced radiation to the longitudinally expanding
fireball. Neither of these features is understood theoretically at
present, but they appear to probe the dynamics of the medium in new
and sensitive ways and must be understood. Three-particle correlations
promise to discriminate clearly among the proposed physics scenarios,
but they are statistically very demanding. High quality multi-hadron
correlation measurements require an order of magnitude more data than
currently available, which can only be achieved with the RHIC II
luminosity and detector upgrades. The large jet yields at the LHC will
enable similar measurements. Detector upgrades and high statistics
datasets will also generate much more detailed investgation of the
``intermediate \pT'' region, where identified particle measurements
(especially correlations) probe the interaction of jets with the
medium.

\ \\
{\bf Fully reconstructed jets:} The inclusive jet spectrum in heavy
ion collisions at the LHC will reach $\ET\sim400$ GeV, while at RHIC
II (following the luminosity upgrade) it will extend beyond
$\ET\sim60$ GeV. At such large jet energies, infrared-safe jet
reconstruction (recovery with good resolution of the full energy of
hard-scattered partons) can be carried out even in the presence of the
large underlying event in heavy ion collisions. Full jet
reconstruction is, per definition, insensitive to details of the
fragmentation. Full jet reconstruction, similar to the $\gamma$/Z+jet
measurements discussed above, will therefore be free of geometric
biases intrinsic to the leading particle analyses currently being
carried out. The full range of modifications of jet structure can
therefore be studied, with qualitatively new observables. The broad
jet energy range will probe the medium over a broad variation in
resolution scale, analogous to study of the \Qsq\ evolution of nucleon
structure in DIS measurements. The jet physics program in heavy ion
collisions at the LHC and RHIC II is still being developed, but such
measurements have the potential to provide deep and qualitatively new
insights into partonic interactions in QCD matter.


\subsubsection{Quarkonium Suppression and Deconfinement}
\label{FutureQuarkonium}

One hallmark of the Quark-Gluon Plasma is {\it deconfinement}, the
dissociation due to color screening of hadronic states that are bound
in vacuum. Twenty years ago, Matsui and Satz proposed that
deconfinement could be observed through strong \JPsi\
suppression. Such suppression has indeed been observed at the SPS by
the NA50 experiment. It is not directly interpretable in terms of
deconfinement, however, since absorption in cold nuclear matter also
contibutes to the observed suppression and its effect must be
disentangled through a systematic study of A+A, p+A, and p+p
collisions. Quarkonium suppression is nevertheless {\it the} essential
signature of deconfinement, and measurement of quarkonium production
in A+A, p+A and p+p collisions is a key element of the RHIC and LHC
heavy ion programs.

\begin{table}
\centering
\begin{tabular}{||c||c|c|c|c||} 
\hline
$q\bar{q}$ & $J/\psi$ & $\chi_c(1P)$ & \PsiPr & $\Upsilon(1S)$ \\ \hline
$T_{dissoc}/T_c$ & 1.7-2.0 & 1.0-1.2 & 1.0-1.2 & $\sim5$ \\ 
\hline
\hline
$q\bar{q}$ & $\chi_b(1P)$ & $\Upsilon(2S)$ & $\chi_b(2P)$ & $\Upsilon(3S)$ \\ \hline
$T_{dissoc}/T_c$ & $\sim1.6$ & $\sim1.4$ & $\sim1.2$ & $\sim1.2$ \\ 
\hline
\end{tabular}
\caption{Dissociation temperatures of various quarkonium states
relative to the deconfinement temperature, from recent finite
temperature lattice calculations in both quenched and
two-flavor QCD.
\label{tab:DissocTemp}}
\end{table}

Recent lattice QCD calculations predict a hierarchy of dissociation
temperatures for different quarkonium states, as shown in
Table~\ref{tab:DissocTemp}. \Chic, \PsiPr, and
\UpsThreeS\ are loosely bound and dissociate near the deconfinement 
transition temperature $T_c$ temperature, while \UpsOneS\ is most
tightly bound and survives well above the transition
temperature. There are still significant theoretical uncertainites in
these estimates, but the general features present in the table suggest
that systematic study of multiple quarkonium states may provide a
powerful {\it differential} probe of color screening and
deconfinement.

Quarkonium production cross sections are generally much smaller than
jet cross sections, and quarkonium measurements at RHIC are only now
coming to maturity. PHENIX has found the same magnitude of \JPsi\
suppression in nuclear collisions at RHIC as was seen by the NA50
collaboration at the SPS, a result which is surprising in light of the
larger energy density measured by jet quenching at RHIC. This may be
due to large opacity of the fireball in both cases, leading to similar
geometric bias and a geometry-driven suppression factor; to a
conspiracy at RHIC of larger initial suppression and significant
production via coalescence at a later stage of the fireball evolution;
or to preferential dissociation of \Chic\ at moderate temperatures
near the phase transition, suppressing its feed-down contribution to
the observed \JPsi\ signal while the \JPsi\ itself survives
(Table~\ref{tab:DissocTemp}). Additional measurements, such as
elliptic flow and the rapidity and transvere momentum dependence of
the suppression, will provide strong constraints on the possible
underlying mechanisms. While the PHENIX measurement has prompted new
theoretical activity to model the various effects, a clear resolution
to this puzzle requires measurements of additional quarkonium states.

The RHIC II luminosity upgrade is required for significant
measurements of all states in the Table except \JPsi. A crucial
experimental test of the relative importance of initial production
vs. coalescence is the collision energy dependence of the suppression,
which can only be studied with the upgraded RHIC luminosity.

\begin{figure}[tbp!]
\centering
\includegraphics[width=0.9\textwidth]{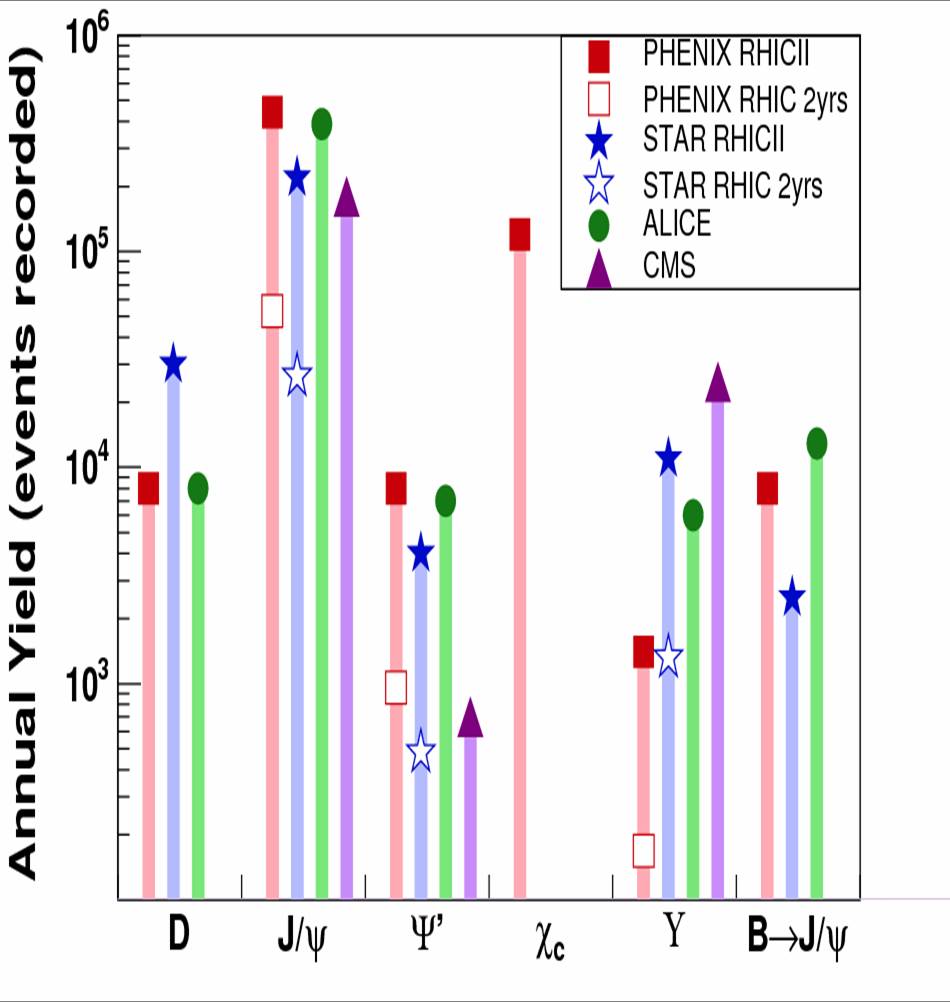}
\caption{Annual yields for heavy flavor production at RHIC II and the LHC.
Open symbols indicate expected total yields for the
near-term RHIC runs in 2007 and 2009, prior to the luminosity upgrade.
\label{fig:HeavyFlavorRates}}
\end{figure}

Figure \ref{fig:HeavyFlavorRates} shows the expected yield of various
heavy flavor states for one year of heavy ion running at RHIC II and
the LHC (annual rates are roughly independent of collision system),
together with expectations for near-term RHIC running prior to the
luminosity upgrade. In all channels except \JPsi, the luminosity
upgrade turns statistically marginal (or worse) measurements into
robust probes of the medium. 

The larger cross sections at the higher LHC energy are approximately
balanced by the increased luminosity and running times at RHIC II, so
that the heavy flavor yields per year are similar. Thus the types and
quality of measurements that can be made at the two facilities will
also be similar. However, there will be a significant difference in
the physics environments at the two facilities that will make the
programs complementary. The higher initial energy density at the LHC
means that the QGP will be created at a significantly higher
temperature. In addition, the factor of ~10 increase in charm pairs
and the factor of ~100 increase in bottom pairs per central collision
at the LHC will have a major impact on the interpretation of heavy
flavor measurements.

At the LHC, all of the charmonium states may be unbound at the highest
temperatures. Thus the prompt charmonium yields at the LHC should be
large and be dominated by coalescence and feed-down
($B\rightarrow\JPsi$), with relatively little contribution from the
primordial \JPsi\ production. Because of its higher binding energy,
bottomonium at the LHC should behave similarly to charmonium at
RHIC. The \UpsOneS\ may remain bound at the highest temperatures at
the LHC while the other bottomonium states will melt. RHIC II and LHC
therefore provide a number of complementary probes of quarkonium
suppression, enabling detailed, differential study of color screening
and deconfinement.

\subsection{Future prospects: summary}

The RHIC community has embarked on a series of integrated upgrades to
the detectors and the accelerator complex, which will provide broad new
capabilities that address the fundamental questions raised by the first
generation of RHIC experiments. Most importantly, these upgrades,
together with significant progress in theory, will provide dramatic
progress in the {\it quantitative} understanding of hot QCD
matter. Progress in this direction has already been made, in
particular in the measurement of the transport parameter
\qhat. Definitive measurements of \qhat, $\eta/s$, and other 
fundamental quantities require the upgrades and progress in theory
discussed in this chapter.

The LHC will soon begin operation, not only with p+p collisions but
also with heavy ion collisions for four weeks per year. The
simultaneous operation of heavy ion experiments at RHIC and LHC offers
an unprecedented opportunity to understand QCD matter in great
depth. Each facility has its strengths, in terms of flexibility in
beams and energies and kinematic reach of hard probes. Equally
important, however, is the ability to carry out the {\it same} jet
quenching or elliptic flow measurements on physical systems evolving
from vastly different initial states. The comparison of RHIC and LHC
measurements promises to give deep insights into the nature of the
experimental probes and their interactions with the medium, and
consequently the hot QCD medium itself.


\newcommand{\ep}{\textit{e}+\textit{p}}
\newcommand{\eA}{\mbox{\textit{e}+A}}
\newcommand{\eN}{\mbox{\textit{e}+N}}
\newcommand{\eAu}{\mbox{\textit{e}+Au}}
\newcommand{\eCa}{\mbox{\textit{e}+Ca}}
\newcommand{\dAu}{\textit{d}+Au}
\newcommand{\pAu}{\textit{p}+Au}
\newcommand{\gev}{\mbox{$\mathrm{GeV}$}}

\section{The Emerging QCD Frontier: The Electron-Ion Collider}
\label{EIC}

Much of the focus in contemporary nuclear physics research is on
mapping and understanding the emergent phenomena from QCD that
determine the unique properties of strongly interacting matter: the
breaking of chiral symmetry that gives light-quark hadrons most of
their mass; the spin, flavor, space and momentum structure of hadrons;
the nearly perfect liquid behavior of the hot matter created in RHIC
collisions; possible color superconductivity in the dense interior of
compact stars.  A key to understanding the rich panoply of QCD
phenomena is identifying conditions under which the theory is amenable
to controlled solution.  Numerical solutions on a space-time lattice
have made impressive advances in the treatment of strongly interacting
matter in equilibrium at both low and high temperatures.  A
perturbative expansion in powers of the running QCD coupling constant
$\alpha_s$ is successful in describing hadron dynamics in high-energy
processes involving large momentum transfer. Interactions of pions and
nucleons at low momentum have been successfully analyzed via chiral
effective field theories.

Recent theoretical advances have introduced a new QCD regime that may
be amenable to a quite different effective field theory approach.
This new interpretability frontier occurs in matter probed at moderate
momentum transfers, where the QCD coupling is still relatively weak,
but at gluon densities high enough to produce extremely strong color
fields that can be treated by classical field theory. This regime is
dominated by direct manifestations of the defining feature of QCD: the
self-interaction of gluons. Gluon splitting and gluon recombination
are predicted to reach a competitive balance, leading to a saturation
of gluon density that should be universal to all strongly interacting
matter probed under suitable conditions.  Hints of this saturation
have been extracted from measurements of electron-proton collisions at
HERA and of deuteron-nucleus and nucleus-nucleus collisions at RHIC.
Saturated gluon densities would have a profound influence on heavy-ion
collisions at the LHC, and may well be the source of certain general
features of high-energy hadron cross sections.  In order to tie these
phenomena together and map the universal properties of gluon-dominated
matter, one needs to probe partonic structure at very low values of
Bjorken $x$, where individual partons carry $<\sim 0.1\%$ of a
nucleon's overall momentum, but within a ``sweet spot" in momentum
transfer ($Q^2$) where the color interaction is neither too weak nor
too strong.

The ideal accelerator to test this classical field theory approach
well into the gluon saturation regime with an \emph{a priori}
understood probe is an Electron-Ion Collider, EIC. Coherent
contributions from many nucleons within a heavy-ion beam particle at
such a collider amplify gluon densities, thereby broadening the $Q^2$
``sweet spot" and extending the effective reach to small $x$-values by
about two orders of magnitude, in comparison with e-p collisions at
the same energy per nucleon.  In addition to providing precocious
entry into the anticipated universal saturation regime, how does the
nuclear environment affect the \emph{path} to saturation? Do the
momentum and space distributions of gluons in nuclei differ in
non-trivial ways from those in nucleons, as has been found for quarks?
Are there small clumps of gluons, or are they more uniformly
distributed? These questions will be addressed by a combination of
deep inelastic inclusive scattering and vector meson production from
nucleons and nuclei.

The addition of \emph{polarized} proton and light-ion beams to collide with polarized
electrons and positrons at EIC would dramatically expand our understanding of the
nucleon's internal wave function.  It would greatly extend the kinematic reach and
precision of deep inelastic scattering measurements of nucleon spin structure. The
contribution of gluons and of sea quarks and antiquarks of different flavor to the
nucleon's spin would be mapped well into the gluon-dominated region.
The study of Generalized Parton Distributions (GPD's) in deep
exclusive reactions will be pushed far beyond presently accessible
energies at JLab, HERA and CERN, extending three-dimensional spatial
maps of the nucleon's internal landscape from the valence quark region
down into the region dominated by sea quarks, antiquarks and gluons.
This extension may be critical for completing the picture of how the
nucleon gets its spin, by providing sensitivity via GPD's to the
orbital motion of sea partons.

High-energy scattering from nucleons in a collider environment lends
itself specifically to study how the creation of matter from energy is
realized in QCD when an essentially massless (and colored) quark or
gluon evolves into massive (and color-neutral) hadrons.  Numerical
solutions of QCD on a space-time lattice cannot provide guidance for
the dynamical process by which the scattered parton picks up other
colored partners from either the QCD vacuum or the debris of the
high-energy collision. Rather, we rely on experiment to map the result
of these parton fragmentation dynamics.  The availability of a
high-energy, high-luminosity polarized electron-ion collider, using
high-efficiency detectors with good particle identification, will
facilitate experiments to measure new features of the fragmentation
process, such as its dependence on quark spin, flavor and motion, and
on passage through nuclear matter.

In short, EIC is a machine that would expand the intellectual horizons
of nuclear physics research into the non-linear heart of QCD, where
gluon self-interactions dominate.  It would address the following
fundamental science questions:

\begin{itemize}

\item[$\bullet$] Does the self-limiting growth of color field strengths in QCD lead to
universal behavior of all nuclear and hadronic matter in the vicinity of these limits?

\item[$\bullet$] How does the nuclear environment affect the distribution of gluons in
momentum and space?

\item[$\bullet$] What is the internal landscape of a nucleon in the region dominated by
sea quarks and gluons?

\item[$\bullet$] How do hadronic final states form from light quarks and massless gluons
in QCD?

\end{itemize}

\noindent It would build on the scientific and technical expertise developed over decades
at the nation's two premier QCD laboratories at Jefferson Lab and
RHIC, but would add new state-of-the-art accelerator technology to
reach its design goals.

In this section, we highlight several of the science programs that EIC
would foster and outline two design options under consideration,
referring the reader to the more detailed White Papers
\cite{eic-wp,eA-wp} that have been written on EIC alone.  We also
describe briefly below the R\&D necessary to demonstrate feasibility
of various aspects of accelerator and detector design for such a
facility.

\subsection{Physics of Strong Color Fields}

{\small \sl With its wide range in energy, nuclear beams, high
luminosity and clean collider environment, the EIC will offer an
unprecedented opportunity for discovery and for the precision study of
a novel universal regime of strong gluon fields in QCD.  The EIC will
allow measurements, in a wide kinematic regime, of the momentum and
spatial distribution of gluons and sea-quarks in nuclei, of the
scattering of fast, compact probes in extended nuclear media, and of
the role of color neutral (Pomeron) excitations in scattering from
nuclei.  These measurements at the EIC will deepen and corroborate our
understanding of the formation and properties of the strongly
interacting Quark Gluon Plasma (QGP) in high energy heavy ion
collisions at RHIC and the LHC.}

\noindent{\bf Strong color fields in nuclei.} One of the major discoveries of the last
decade was just how dominant a role gluons play in the wave function of a proton viewed by
a high-energy probe with high spatial resolution (\emph{i.e.}, with large 4-momentum
transfer squared $Q^2$).  HERA deep inelastic scattering data revealed that the density of
partons, especially gluons, in the plane transverse to the probe momentum grows rapidly
with decreasing parton momentum fraction $x$.  This growth is attributable in QCD to the
successive emission of soft partons by higher-momentum partons. The resulting gluon field
can be treated linearly within QCD when $x$ and $Q^2$ are not too small.  But for given
$x$, the dynamics of the gluon fields becomes highly non-linear below a certain saturation
momentum scale $Q_s^2$.  At low $x$, where parton densities are quite high, the
recombination of soft gluons into harder ones sets in as the leading non-linear
interaction to tame further growth of the parton densities. If the saturation momentum is
large on a typical QCD scale, $Q_s \gg \Lambda_{\rm QCD}$, then the coupling strength
$\alpha_s(Q_s^2) \ll 1$ and the gluon dynamics can be described with weak-coupling
techniques. The occupation number of gluon field modes with transverse momenta below $Q_s$
saturates at values $\sim 1/\alpha_s(Q_s^2) \gg 1$, so that the probe sees a very strong,
essentially classical, color field frozen by time dilation, a system often referred to as
the "color glass condensate" (CGC).  A goal of theoretical treatments of this high-density
QCD matter is to establish a rigorous effective field theory approach for controlled
inclusion of higher-order effects beyond the CGC limit.

\begin{figure}[thp]
\vskip 0.1cm
\centering
\includegraphics[width=0.7\columnwidth]{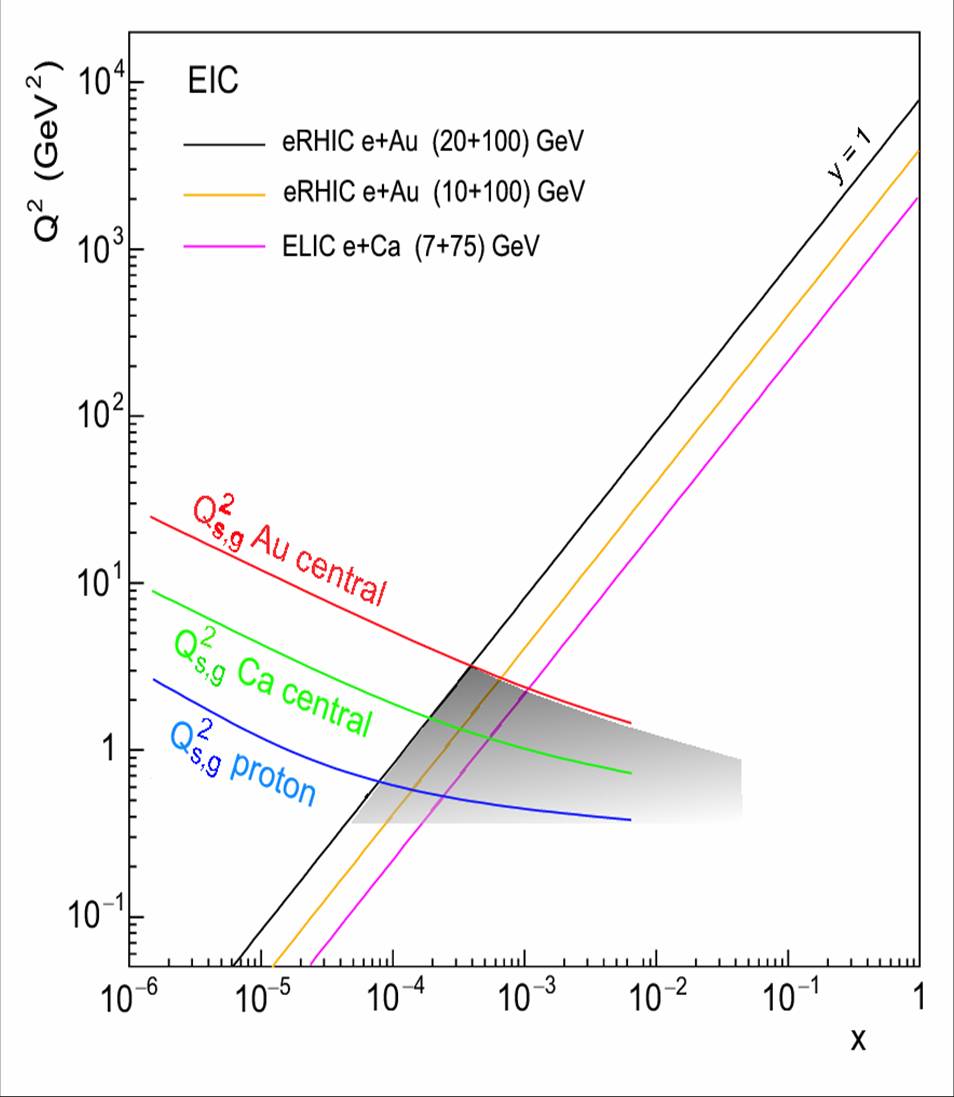}
\caption{Kinematic acceptance and exposure of the predicted gluon saturation regime in the
($x,Q^2$) plane for the EIC. The accessible regions fall to the right of the three
diagonal straight lines, representing different choices for beam energies (per nucleon in
the case of ion beams) and maximum mass of the ion beams.  Curves showing the gluon
saturation scale $Q_s^2$ for protons and for central collisions with Ca and Au nuclei are
superposed on the kinematic acceptance. The shaded area indicates the kinematically
accessible region of saturated gluon density that should be reached in the maximum-energy
e+Au collisions considered. \label{fig:xq2}}
\end{figure}

Since the saturation momentum grows slowly with decreasing $x$ (see
Fig.~\ref{fig:xq2}), so does the window ($\Lambda_{\rm QCD} \ll Q \ll
Q_s$) into the CGC regime.  However, a much more effective opening of
this window can be arranged by exploiting the Lorentz contraction of a
fast-moving nucleus, which amplifies the parton density in proportion
to the nuclear diameter, so that $Q_s^2 \propto A^{1/3}$. Thus, as
illustrated in Fig.~\ref{fig:xq2}, one can enter the predicted
saturation regime in e-Au collisions at $x$-values a couple of orders
of magnitude larger than what would be required in e-p collisions at
the same $Q^2$.  An electron-ion collider thus represents the most
robust and cost-effective approach to study the physics of these
strong color fields.  Can a clear saturation scale be identified
experimentally?  Are the properties of partonic matter in the
saturation regime indeed {\it{universal}} to all hadrons and nuclei?
Are these properties consistent with inferences from particle
multiplicities and momentum spectra observed at RHIC and with dynamics
soon to be explored in heavy-ion collisions at the LHC?  Can the
properties of saturated gluon fields in heavy nuclei provide a natural
explanation for the very rapid thermalization inferred from analysis
of relativistic heavy-ion collisions? These questions will be
addressed via deep inelastic scattering (DIS) and other cleanly
interpretable electromagnetic processes at EIC, as explained in more
detail below.

\noindent{\bf Measurements of momentum distributions of gluons and sea quarks in nuclei.}
Gluon momentum distributions overwhelm their quark counterparts in the proton for $x<\sim 0.01$. DIS experiments have established that quark and gluon distributions in
nuclei exhibit ``shadowing'': they are modified significantly relative
to their distributions in the \emph{nucleon} wavefunction. However,
the detailed nature of gluon shadowing at $x<\sim 0.01$ is {\it
terra incognita} in QCD. This physics, bearing directly on the
universality of gluon saturation, can be fully studied in
electron--nucleus scattering at the EIC, over the broad kinematic
coverage shown in Fig.~\ref{fig:xq2}.

\begin{figure}[tbp!]
\centering
\includegraphics[width=0.8\columnwidth]{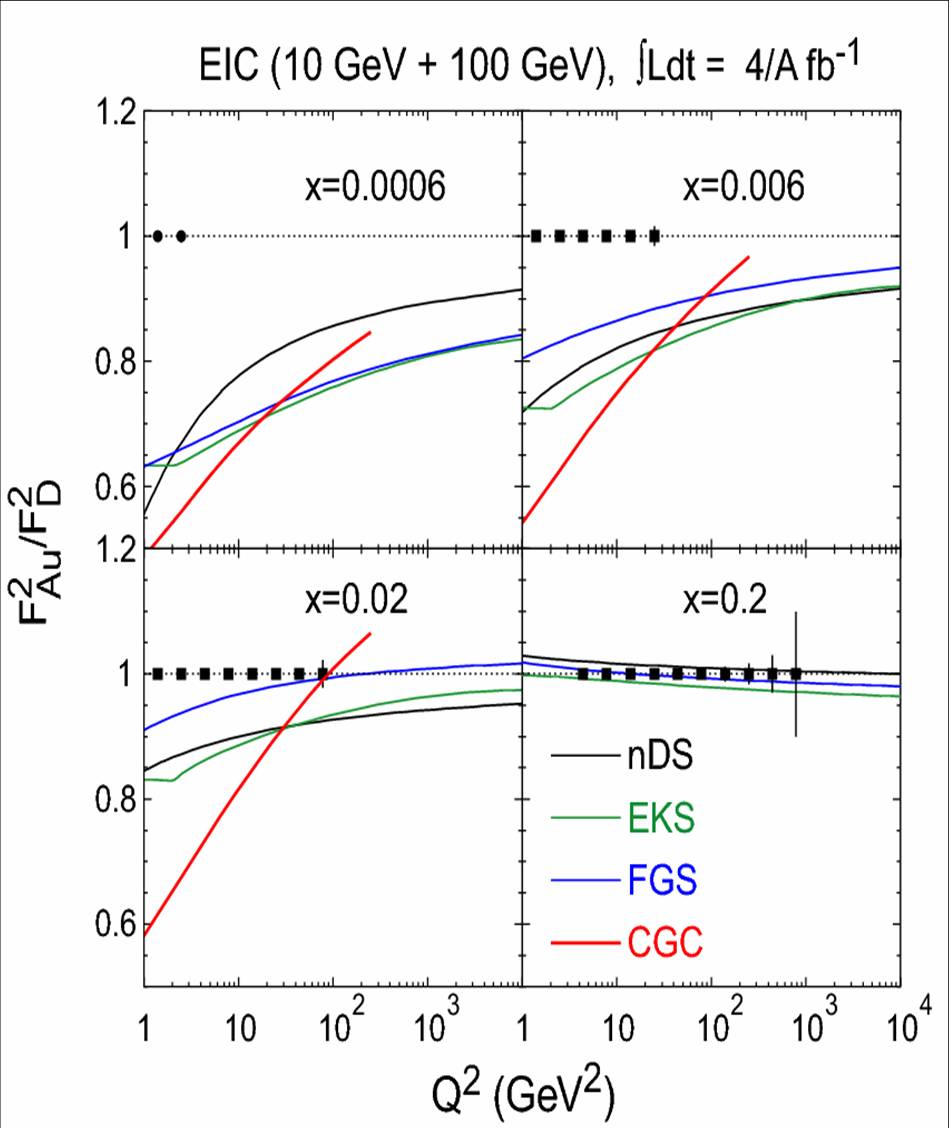}
\caption{The ratio of the structure function $F_2^{\rm Au}$ in
Au nuclei relative to the structure function $F_2^{\rm D}$ in
deuterium nuclei as a function of $Q^2$ for several bins in $x$.  The
filled circles and error bars correspond respectively to the estimated
kinematic reach in $F_2$ and the statistical uncertainties for a
luminosity of 4/A\,fb$^{-1}$ with the EIC.  The curves labeled nDS,
EKS and FGS correspond to different parameterizations of parton
distributions at the initial scale for pQCD evolution, while the one
labeled CGC corresponds to a Color Glass Condensate model prediction
applicable at small $x$.
\label{fig:F2-4}}
\end{figure}

The inclusive DIS structure functions $F_2^A(x,Q^2)$ and $F_L^A(x,Q^2)$ offer the most
precise determination of quark and gluon momentum distributions in nuclei. Independent
extraction of $F_2^A$ and $F_L^A$ is only possible via measurements over a range of center
of mass energies, an essential requirement of the EIC. The $F_2^A$ structure function is
directly sensitive to the sum of quark and anti-quark momentum distributions in the
nucleus; at small x, these are predominantly sea quarks. Information on the gluon
distribution in the nucleus, $G^A(x,Q^2)$, can be indirectly garnered from the well-known
logarithmic scaling violations of $F_2^A$ with $Q^2$, $\partial F_2^A/\partial \ln(Q^2)$.
In Fig.~\ref{fig:F2-4} we show projections for the normalized ratio of $F_2^A(x,Q^2)$ in
gold relative to deuterium from a saturation (CGC) model in comparison to the usual linear
evolution of perturbative QCD for three models incorporating differing amounts of
shadowing. Saturation of gluon densities in the CGC model is manifested by the weak $x$-
and $Q^2$-dependence of the slope $\partial F_2^{Au}/\partial \ln(Q^2)$ at low $x$ and
moderate $Q^2$. The projected statistical precisions attainable for inclusive DIS
measurements with 10\,GeV electrons on 100\,GeV/nucleon Au nuclei and an integrated
luminosity of 4/A\,fb$^{-1}$, also shown in Fig.~\ref{fig:F2-4}, suggest that EIC data can
readily distinguish among differing model predictions.

The structure function $F_L^A \equiv F_2^A - 2xF_1^A$ for absorption of longitudinal
photons by the proton vanishes in the naive parton model, but in QCD it is proportional at
small $x$ to the gluon momentum distribution. Hence, its measurement will allow a new and
independent direct determination of $G^A(x,Q^2)$ in the low-$x$ region where little is
presently known. The high precision attainable for both $F_2$ and $F_L$ at EIC will
facilitate definitive tests of the universality of saturated gluonic matter.  Measurements
for different nuclei, $x$ and $Q^2$ values can be combined in a single plot of the
structure functions vs. $Q^2 x^\gamma /A^\delta$ to search for values of the adjustable
powers $\gamma$ and $\delta$ that yield a universal curve, and hence define the $x$- and
$A$-dependence of the saturation scale $Q_s^2(x,A)$.

Additional strong sensitivity to gluon densities in nuclei will be provided by
semi-inclusive and exclusive final states. An example of the former is di-jet production
in e-A collisions, which is dominated at EIC energies by the photon-gluon fusion process.
An exclusive example is elastic vector meson production $\eA \longrightarrow (\rho,\phi,
J/\psi)$+A, where forward cross sections for longitudinal virtual photons depend on the
square of the gluon density.

\noindent{\bf The gluon spatial distribution.} The spatial distribution of gluons in a
nucleus provides a complementary handle on the physics of strong color
fields and has important ramifications for a wide range of final
states in hadronic and nuclear collisions. Information on the spatial
distribution can be inferred from forward vector meson production in
e-A, which can be viewed at small $x$ as the result of coherent
interactions of quark-antiquark fluctuations of the virtual photon
with the nucleus. The differential cross section for the vector
mesons, as a function of momentum transfer $t$ along the proton line,
can be analyzed to extract a survival probability of these small color
dipole fluctuations as a function of impact parameter $b$ at which the
dipole traverses the nucleus.  The survival probability is, in turn,
sensitive to the strength of the gluon field seen.  Systematic studies
of vector meson production over a wide range of kinematic conditions
and for several ion species can thereby illuminate the $b$-dependence
as well as the $A$-dependence of the saturation scale.

\noindent{\bf Color neutral (Pomeron) excitations in scattering off nuclei.} Another
predicted manifestation of strong gluon fields in QCD is an enhanced
probability for a high-energy probe to interact with a color-neutral
multi-gluon excitation of the vacuum -- an excitation that may be
associated with the so-called Pomeron -- leaving the target nucleus
intact. These interactions lead to diffractive final states that may
dominate forward scattering.  At HERA, an unexpected discovery was
that diffraction accounted for ~15\% of the total \ep\
cross-section. This is a striking result implying that a proton at
rest remains intact one seventh of the time when struck by a 25\,TeV
electron. The effect may be even more dramatic in nuclei. Several
models of strong gluon fields in nuclei suggest that large nuclei will
remain intact nearly 40\% of the time in EIC collisions, in comparison
to the quantum mechanical black disk limit of 50\%. Measurements of
coherent diffractive scattering on nuclei are easier in the collider
environment of EIC than in fixed-target experiments, but nonetheless
place strong demands on the forward acceptance of detectors.  With
suitable detectors, EIC measurements should be able to distinguish the
onset of non-linear dynamics for the gluon field, leading to a weak
$x$-dependence but strong $Q^2$-dependence of the ratio of diffractive
structure functions for heavy
\emph{vs.} light nuclei. These dependences are distinct from those expected in
non--perturbative (``soft'' Pomeron) models of diffractive scattering.

\noindent{\bf Fast probes of an extended gluonic medium.} How are the propagation of fast
partons and their space-time evolution into hadrons affected by traversal of nuclear
matter characterized by strong gluonic fields? Semi-inclusive DIS (SIDIS) experiments at
EIC, with high-momentum hadrons detected in coincidence with scattered electrons for a
wide range of kinematic conditions and ion species, will use nuclei as femtometer-scale
detectors to study these issues in cold nuclear matter. These experiments will provide an
essential complement to studies of jet quenching in the hot matter produced in RHIC
heavy-ion collisions.  The RHIC jet quenching studies have produced a series of striking
and surprising results: a strong suppression of high-momentum hadrons usually attributed
to rapid energy loss of partons traversing matter of high color charge density, but little
apparent dependence of the suppression factor on quark flavor, in sharp contrast to
expectations from perturbative QCD models of the parton energy degradation. SIDIS on
\emph{fixed} nuclear targets has so far revealed an analogous but weaker suppression of
light hadron production in cold nuclear matter.  EIC will enormously expand the virtual
photon energy range in such studies, from 2--25 GeV in the HERMES experiment at HERA to
$10\,\gev\ < \nu < 1600\,\gev$, thereby providing access to the kinematic region relevant
for LHC heavy-ion collisions and to such important new issues as the suppression of
heavy-flavor mesons travelling through cold nuclear matter.

One of the basic physics questions to be answered here concerns the
time scale on which the color of the struck quark is neutralized,
acquiring a large inelastic cross-section for interaction with the
medium. The parton energy loss models used to interpret RHIC results
assume long color neutralization times, with ``pre-hadron'' formation
outside the medium and quark/gluon energy loss as the primary
mechanism for hadron suppression.  Alternative models assume short
color neutralization times with in-medium ``pre-hadron'' formation and
absorption as the primary mechanism. There do exist hints of short
formation times from HERMES data and JLab preliminary data, but these
must be pursued over the wider kinematic range and much broader array
of final-state channels that can be explored at EIC.

\subsection{A New Era of Hadronic Physics \label{hadron}}

{\small \sl The EIC will provide definitive answers to compelling
physics questions essential for understanding the fundamental
structure of hadronic matter. It will allow precise and detailed
studies of the nucleon in the regime where its structure is
overwhelmingly due to gluons and to sea quarks and anti-quarks. Some
of the scientific highlights at the EIC in this area would be: (1)
definitive answers to the question of how the proton's spin is carried
by its constituents, (2) determination of the three-dimensional
spatial quark and gluon structure of the proton, (3) precision study
of the proton's gluon distribution over a wide range of momentum
fractions, and (4) maps of new spin-dependent features of the quark
fragmentation process. In the following we briefly address three of
these highlights of future research in hadronic physics.}

\begin{figure*}[tbp]
\centering
\includegraphics[width=0.85\columnwidth]{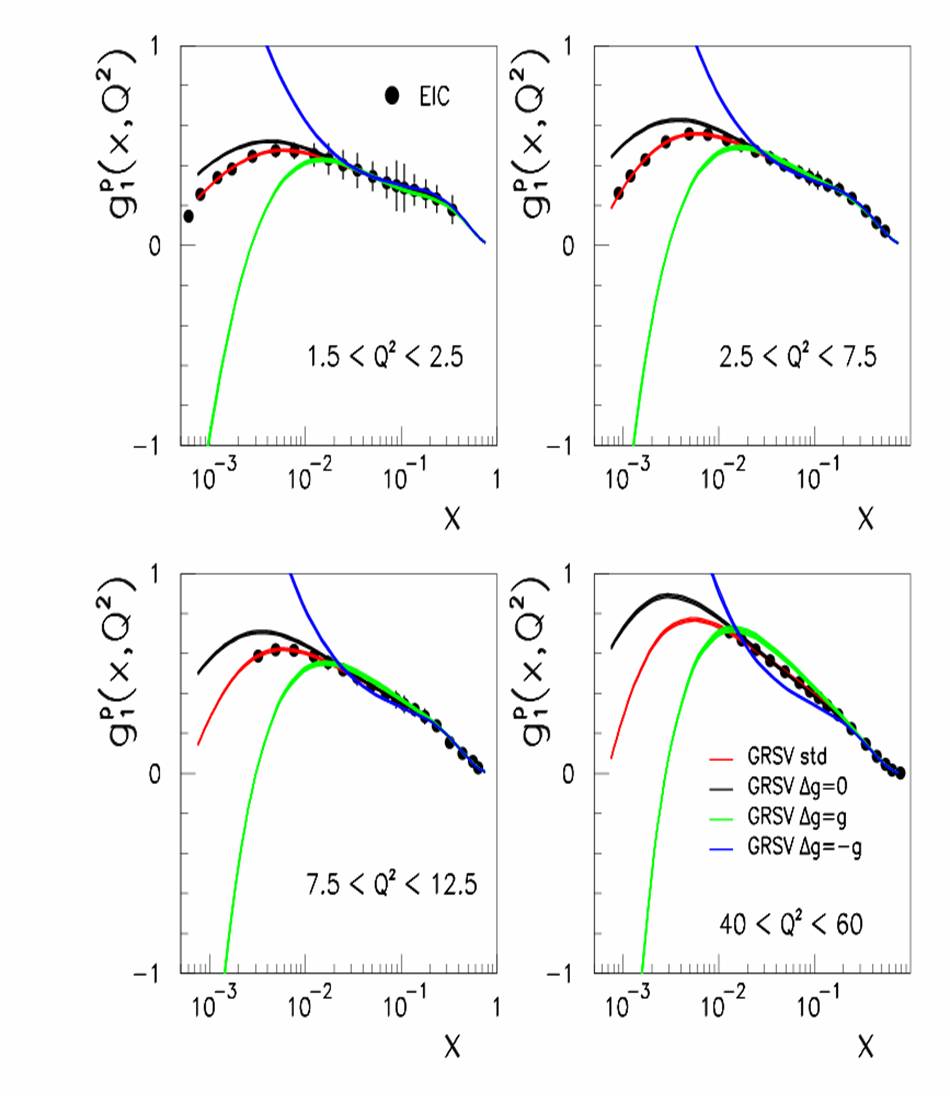}
\caption{Projected EIC data for the
proton structure function $g_1(x,Q^2)$ as a function of $x$ in four
$Q^2$ bins, for 7~GeV electrons colliding with 150~GeV protons at an
integrated luminosity of $5\,{\mathrm{fb}}^{-1}$. The curves show
theoretical predictions based on different sets of spin-dependent
parton distribution functions that mostly differ in the gluon helicity
distribution.
\label{g1scval}}
\end{figure*}

\noindent {\bf The spin structure of the proton.} Few discoveries in nucleon structure
have had a bigger impact than the surprising finding that quarks and
anti-quarks together carry only about a quarter of the nucleon's
spin. Determining the partonic source of the ``missing" spin in this
complex composite system has developed into a world-wide quest central
to nuclear physics. The sum rule $$
\frac{1}{2}=\frac{1}{2}\Delta \Sigma + L_q + \Delta G + L_g  \;
$$
states that the proton spin projection along its momentum is the sum of the quark and
gluon intrinsic spin ($\Delta \Sigma$, $\Delta G$) and orbital angular momentum ($L_q$,
$L_g$) contributions. EIC with its unique high luminosity, highly polarized electron and
nucleon capabilities, and its extensive range in center-of-mass energy, will allow DIS
access to quark and gluon spin contributions at substantially lower momentum fractions $x$
than important current and forthcoming experiments at RHIC, DESY, CERN and JLab.  A key
measurement at the EIC would be of the spin-dependent proton structure function
$g_1(x,Q^2)$ of the proton over a wide range in $Q^2$, and down to $x\sim 10^{-4}$.
Studies of the scaling violations of $g_1(x,Q^2)$ prove to be a most powerful and clean
tool to determine the spin contribution by gluons. This is demonstrated by
Fig.~\ref{g1scval}, which shows projections for EIC measurements of $g_1(x,Q^2)$ in
comparison with four model predictions that make different assumptions regarding the sign
and magnitude of the gluon spin contribution to the proton spin.  Each of these models is
compatible with the currently available polarized fixed-target DIS data.  While data from
polarized proton collisions at RHIC are already beginning to establish preferences among
these particular four models at $x>\sim0.01$, the RHIC data will not be able to
constrain the shape of the gluon helicity distribution at lower $x$, where the density of
gluons rapidly increases. The great power of the EIC in providing precise information on
$\Delta G(x<\sim0.01)$ is evident.

\newpage


With polarized $^3$He beams at an EIC, measurements of $g_1$ would also be possible off
polarized neutrons, allowing a precision test of the fundamental Bjorken sum rule, which
relates the proton and neutron spin structure via the axial weak coupling strength
measured in neutron beta-decay. Furthermore, semi-inclusive DIS measurements, for which a
specific hadron is detected from the struck quark jet, would provide information with
unprecedented detail on the individual contributions by quark and anti-quark spins to the
proton spin, testing models of nucleon structure and lattice QCD calculations.

There are various avenues for investigating the role of orbital angular momenta in nucleon
structure. One of them is the study of correlations of the transverse momentum of a parton
in the nucleon with the nucleon spin transverse to its momentum. Such correlations produce
characteristic patterns of azimuthal-angular dependences for final-state hadrons in SIDIS
experiments. Initial experimental results from fixed-target SIDIS indicate the presence of
such correlations. Measurements at an EIC would allow precision studies of such orbital
effects. An alternative approach will utilize deep exclusive reactions to extract
generalized parton distributions (GPDs), to which we turn next.  The GPDs provide unique
access to the total -- spin plus orbital -- angular momentum contributions of quarks and
gluons, as well as to many other important aspects of nucleon structure.  While initial
maps of GPDs in the valence-quark region will be carried out with the 12 GeV upgrade at
JLab, access to orbital contributions associated with virtual mesons in the nucleon wave
function will require the EIC kinematic reach well into the region of the quark-antiquark
sea.

\noindent {\bf Measurements of Generalized Parton Distributions.} GPDs may be viewed as
the Wigner quantum phase space distributions of the nucleon's constituents -- functions
describing the simultaneous distribution of particles with respect to position and
momentum in a quantum-mechanical system, representing the closest analog to a classical
phase space density allowed by the uncertainty principle. In addition to information about
spatial density (form factors) and momentum density (parton distribution), these functions
describe correlations of the two, i.e., how the spatial shape of the nucleon changes when
one probes quarks and gluons of different wavelengths. The concept of GPDs has
revolutionized the way scientists visualize nucleon structure, in the form of either
two-dimensional tomographic images (analogous to CT scans in medical imaging) or genuinely
six-dimensional phase space images. In addition, GPDs allow us to quantify how the angular
momenta of partons in the nucleon contribute to the nucleon spin.

Measurements of GPDs are possible in hard exclusive processes such as deeply virtual
Compton Scattering (DVCS), $\gamma^\ast p\to \gamma p$. The experimental study of these
processes is typically much more challenging than of traditional inclusive DIS. In
addition to requiring substantially higher luminosities (because of small cross sections)
and the need for differential measurements, the detectors and the interaction region have
to be designed to permit full reconstruction of the final state. 

\begin{figure}[tp]
\vskip -0.15cm
\centering
\includegraphics[width=0.75\columnwidth]{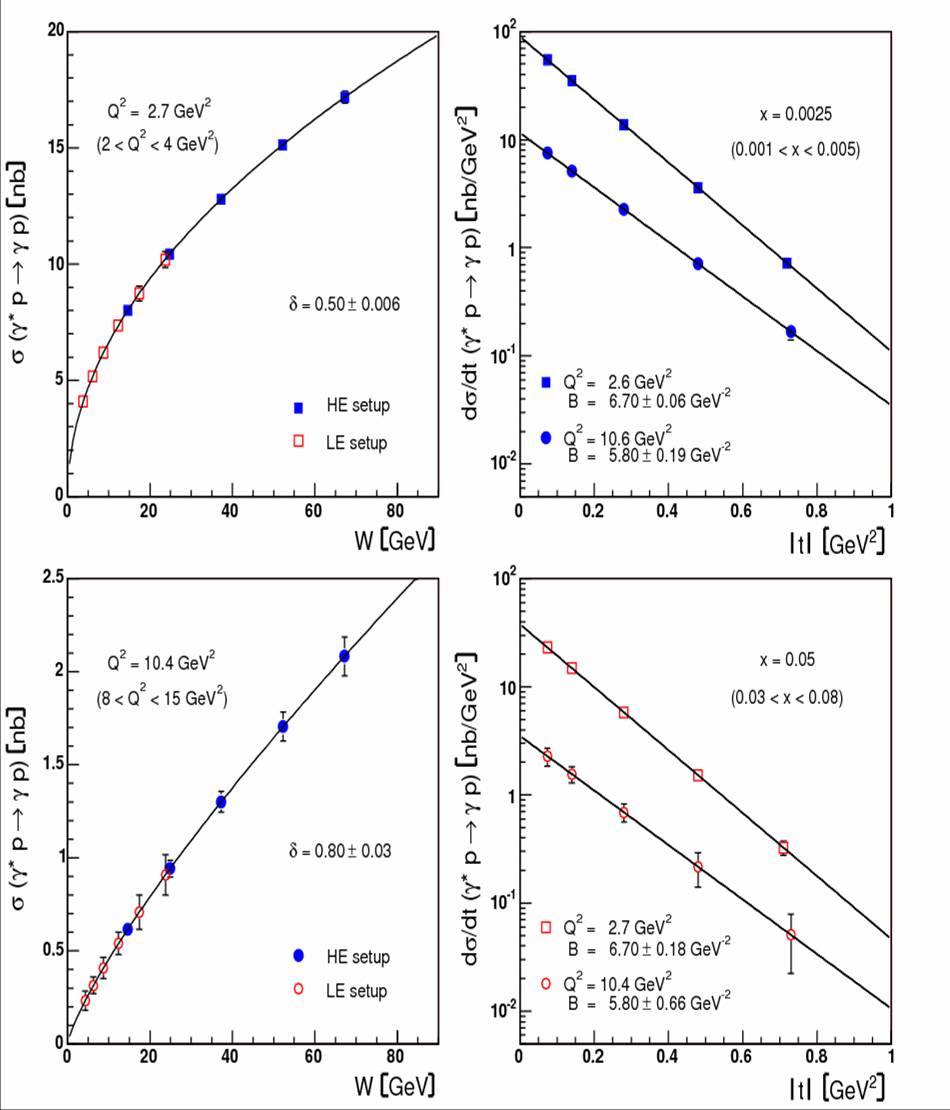}
\caption{Left: projected results for total DVCS cross section measurements with an EIC, as
a function of invariant $\gamma^\ast p$ mass $W$, for two values of $Q^2$. Right: $t$
differential DVCS cross section for two representative values of $x$ and $Q^2$. The
projections assume a high-energy setup (10~GeV on 250~GeV), with an integrated luminosity
of $530\,{\mathrm{pb}}^{-1}$ for the smaller $x$-value, and a low-energy setup (5~GeV on
50~GeV) with $180\,{\mathrm{pb}}^{-1}$ for the larger $x$-value. The estimates of the
event rates here assume 100\% detector acceptance. \label{dvcs}}
\vskip -0.35cm
\end{figure}

A properly designed collider is much better suited for this purpose than a fixed-target experiment. A collider
also achieves momentum transfers of the order $Q^2\sim 10$~GeV$^2$, where higher-twist QCD
corrections in the GPD analysis are under control. The EIC would allow unique access to
the gluon and sea-quark and anti-quark GPDs, entirely complementary to what will be
achieved by the 12-GeV upgrade program at JLab. This would be possible through study of a
variety of exclusive final states, ranging from photons to pions, kaons and $J/\psi$. As
an example of the potential of an EIC in this area, we show in Fig.~\ref{dvcs} the
expected uncertainties of measurements of the DVCS cross section. In particular, we show
the cross section differential in $t$, the momentum transfer on the nucleon line. By a
Fourier transform, the $t$-dependence encodes the information about the transverse spatial
distribution of partons in the proton. One can see that excellent statistics can be
obtained in fully differential measurements in $x$, $Q^2$ and $t$, and over a wide
kinematic range. This will allow for precise extraction of information about the nucleon
GPDs and for numerous detailed studies, for example, of their $Q^2$-evolution.


\noindent \textbf{Spin-dependent Quark Fragmentation.} Semi-inclusive DIS experiments at a
high-luminosity polarized EIC will map the spin-dependence of the process by which quarks
transform to jets of hadrons.  Recoiling quarks from a polarized proton will initiate the
fragmentation process with a spin orientation preference.  How does this preference affect
the yields, momenta and spin preferences of various types of hadronic fragments, and what
do such effects teach us about the fragmentation dynamics? It is already apparent from
measurements in electron-positron collisions and in fixed-target SIDIS that there are
correlations between the momentum components of hadron fragments transverse to the jet
axis and any quark spin preference transverse to its momentum.  In addition to systematic
exploration of these initial hints at EIC, it may be possible for selected final-state
hadrons -- e.g., $\rho$-mesons -- reconstructed from their decay daughters to correlate
their density matrices with the spin orientation of the fragmenting quark.  In combination
with the study of in-medium fragmentation in e-A collisions at EIC, such measurements are
likely to launch a new stage in modeling how quarks accrete colored partners from the
vacuum or their environment to form colorless hadrons.

\newpage

\subsection{Accelerator Designs}

{\small \sl A high luminosity (at or above 10$^{33}$ cm$^{-2}$s$^{-1}$) Electron-Ion
Collider, covering the full range of nuclear masses $A$ with variable center-of-mass
energy in the range of 20 to 100 GeV/nucleon, and the additional capability of colliding
polarized protons and light-ions with polarized electrons and positrons, appears to be the
ideal accelerator to explore these fundamental questions of QCD and expand nuclear physics
research into the gluon-dominated regime. Presently there are two distinct design
approaches to an EIC: eRHIC, based on the RHIC ion complex, and ELIC, using CEBAF as a
full energy injector into an electron storage ring. Research and development needed for a
detailed design of each approach is outlined in this section.}\\

\begin{figure}[tbp!]
\vskip -1.cm
\centering
\includegraphics[width=0.9\textwidth]{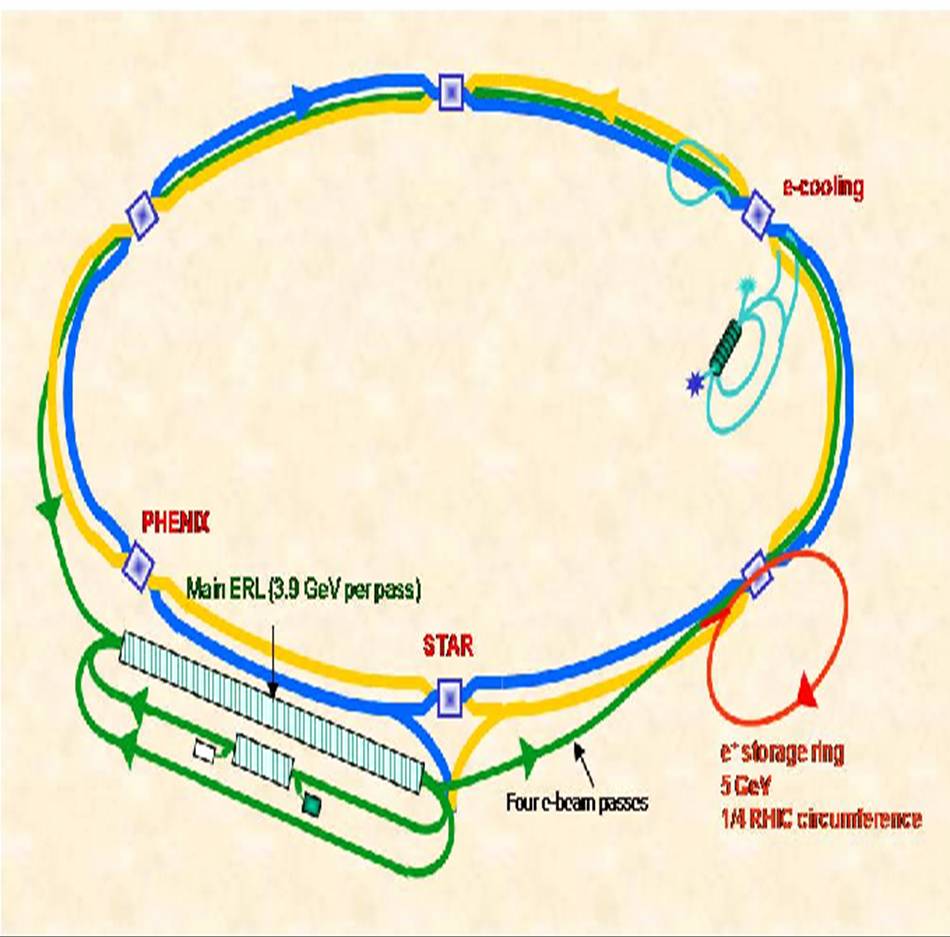}
\vskip 0.5cm
\includegraphics[width=0.9\textwidth]{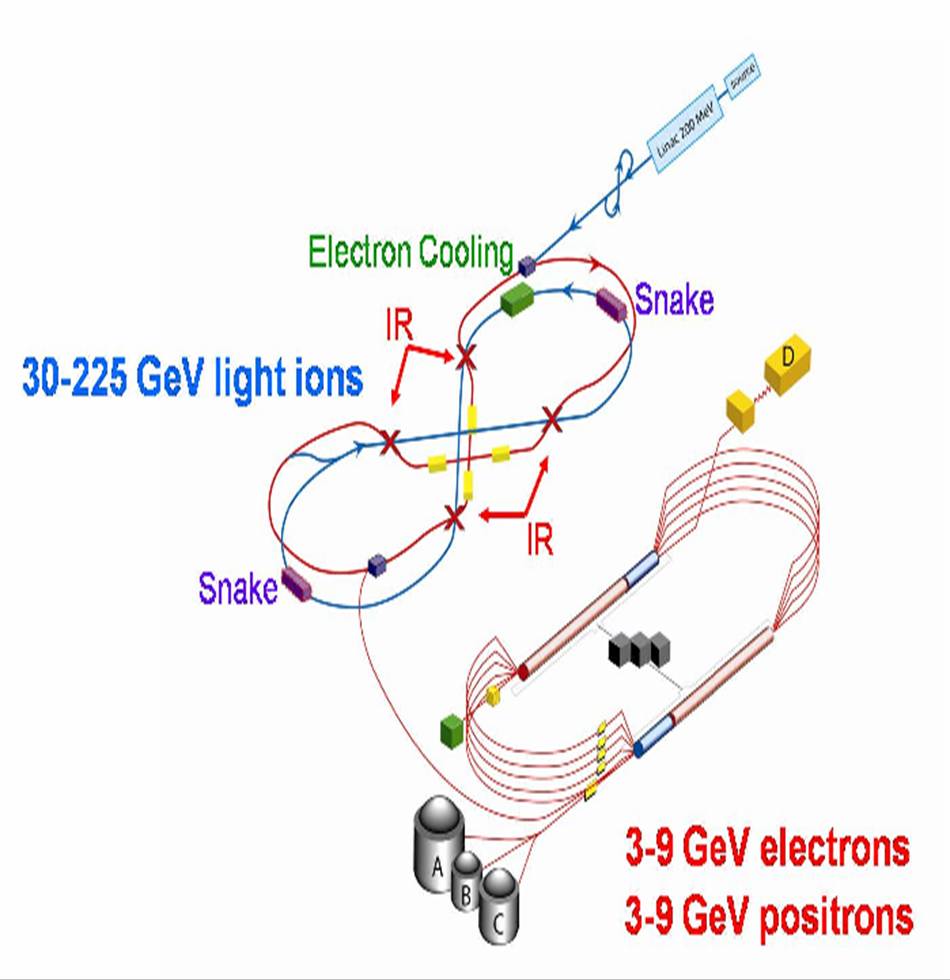}
\vskip 0.75cm
\caption{Design layouts of the ERL-based eRHIC, and the CEBAF-based ELIC colliders.
\label{fig:EIC}}
\end{figure}

\noindent{\bf eRHIC} Two accelerator design options for eRHIC were developed in parallel
and presented in detail in the 2004 Zeroth-Order Design Report\cite{erhic}. Presently the
most promising option is based on the addition of a superconducting Energy Recovery Linac
(ERL) to provide the polarized electron beam. This ERL-based design option can achieve
peak luminosity of 2.6 $\times$ 10$^{33}$ cm$^{-2}$s$^{-1}$ for e-p collisions, with the
potential for improvement. The peak luminosity per nucleon for electron-Au collisions is
2.9 $\times$ 10$^{33}$ cm$^{-2}$s$^{-1}$ for 100 GeV/N gold ions colliding with 20 GeV
electrons. R\&D for a high-current polarized electron source and high-energy and
high-current ERL are needed to achieve these design goals. A second option is based on the
addition of an electron storage ring to provide polarized electron or positron beams. This
option is technologically more mature and promises peak e-p luminosity of 0.47 $\times$
10$^{33}$ cm$^{-2}$s$^{-1}$. The general layout of the ERL-based design option of the
eRHIC collider is shown in Fig.~\ref{fig:EIC}. A polarized electron beam is generated in a
photo-injector and accelerated to the energy of the experiment in the ERL. After colliding
with the hadron beam in as many as four detector locations, the electron beam is
decelerated to an energy of a few MeV and dumped. Positron beam is possible with the
addition of a conversion system and a compact storage ring, at one quarter of the RHIC
circumference, for positron accumulation, storage and self-polarization. In the present
design, the ERL provides electrons in the energy range from 3 to 20 GeV, leading to a
center-of-mass energy range from 25 to 140 GeV in combination with RHIC proton beams.\\

\newpage
The main highlights of the ERL-based eRHIC design are:
\begin{itemize}
\item{luminosity of 10$^{33}$ cm$^{-2}$s$^{-1}$ and higher in electron-hadron collisions}
\item{high electron beam polarization ($\sim$80\%)} \item{full polarization transparency
at all energies for the electron beam} \item{multiple electron-hadron interaction points
(IPs) and detectors} \item{$\pm$3m ``element-free'' straight section(s) for detector(s)}
\item{ability to take full advantage of electron cooling of the hadron beams} \item{easy
variation of the electron bunch frequency to match it with the ion bunch frequency at
different ion energies}
\end{itemize}

\noindent{\bf ELIC} ELIC is an electron-ion collider with center of mass energy of 20 to
90 GeV and luminosity up to 8 $\times$ 10$^{34}$ cm$^{-2}$s$^{-1}$ (at a collision
frequency of 1500 MHz). It is described in detail in the 2007 Zeroth Order Design Report
\cite{elic} and shown schematically in Fig.~\ref{fig:EIC}. This high-luminosity collider
is envisioned as a future upgrade of CEBAF, beyond the 12 GeV Upgrade, and compatible with
simultaneous operation of the 12 GeV CEBAF (or a potential extension to 24 GeV) for
fixed-target experiments.  The CEBAF accelerator with polarized injector is used as a
full-energy injector into a 3-9 GeV electron storage ring. A positron source is envisioned
as an addition to the CEBAF injector for generating positrons that can be accelerated in
CEBAF, accumulated and polarized in the electron storage ring, and collide with ions with
luminosity similar to the electron-ion collisions. The ELIC facility is designed for a
variety of polarized light ion species: p, d, $^3$He and Li, and unpolarized light to
heavy (up to A $\sim$ 200) ion species. To attain the required ion beams, an ion facility
must be constructed, a major component of which is a 30-225 GeV collider ring located in
the same tunnel and below the electron storage ring. A critical component of the ion
complex is an ERL-based continuous electron cooling facility, anticipated to provide low
emittance and simultaneously very short ion bunches. ELIC is designed to accommodate up to
four intersection points (IP's), consistent with realistic detector designs. Longitudinal
polarization is guaranteed for protons, electrons, and positrons in all four IP's
simultaneously and for deuterons in up to two IP's simultaneously.

An alternate design approach for ELIC is based on the linac-ring concept, in which CEBAF
operates as a single-pass ERL providing full energy electrons for collisions with the
ions. Although this approach promises potentially higher luminosity than the ring-ring
option, it requires significant technological advances and associated R\&D. The main
highlights of the ELIC design are:
\begin{itemize}
\item{``Figure-8'' ion and lepton storage rings ensure spin preservation and ease of spin
manipulation} \item{spin transparency to energy for all species} \item{unprecedented
luminosity at the 10$^{35}$ cm$^{-2}$s$^{-1}$ level} \item{four interaction regions with
$\pm$2m element-free region} \item{the present JLab DC polarized electron gun routinely
delivers $\sim$85\% polarization and meets the beam current requirements for filling the
storage ring} \item{the 12 GeV CEBAF accelerator can serve as an injector to the ring}
\item{collider operation remains compatible with 12 GeV CEBAF operation for a fixed-target
program}
\end{itemize}

\vskip 0.25cm

\noindent{\bf R\&D Required}\\

{\sl I. Common R\&D Topics} In order for either eRHIC or ELIC to reach luminosity at or
above 10$^{33}$ cm$^{-2}$s$^{-1}$ level, R\&D on high energy electron cooling and on the
production of polarized $^3$He beams is required. Electron cooling is required to achieve
the design transverse emittances, to counteract the effects of intrabeam scattering, and
in the case of ELIC to reach short ion bunches. An electron cooling system based on ERL
technology is presently under development for RHIC-II, intended to lead to an order of
magnitude higher ion-ion luminosities in RHIC. The same system will be used for eRHIC.
$^3$He ions have not yet been used for experiments. EBIS, the new ion source under
construction at BNL, will provide the ability to produce polarized $^3$He beams, given a
$^3$He source. In addition, R\&D will be required on a variety of detector and polarimetry
items, such as the development of cost-effective and compact high-rate tracking and
associated readout systems, small angle detector instrumentations, multi-level trigger
systems and precision ion polarimetry.\\

{\sl II. R\&D Required for eRHIC} R\&D applicable to both ERL and ring-ring options for
eRHIC is required in order to increase the number of bunches in RHIC from 111 to 166, and
for better understanding of the machine tolerances required for $^3$He polarization
preservation in RHIC and its injectors. In addition, the ERL eRHIC design requires R\&D on
high-current polarized electron sources and on high-energy and high-current energy
recovery. To achieve the design eRHIC luminosities, 260 mA average current is required
from a polarized electron source. The best existing source, at JLab's CEBAF accelerator,
operates at approximately 0.3 mA of average current (1 mA is expected to be reached
shortly) with current densities of about 50 mA/cm$^2$. The development of large cathode
guns should provide a path to electron currents of tens to hundreds of milliamps. The
eRHIC ERL is envisioned to employ state-of-the-art 703.75 MHz 5-cell SRF cavities. The
cavity design was developed at BNL in the course of the electron cooling project and
allows the minimization and efficient damping of the higher-order modes, opening a way for
higher electron currents. Simulations of multi-bunch and multi-pass breakup instabilities
showed that the design eRHIC currents can be achieved in an ERL based on this cavity.\\

{\sl III. R\&D Required for ELIC} With the exception of electron cooling, no additional
R\&D is necessary for ELIC at the luminosity level of 10$^{33}$ cm$^{-2}$s$^{-1}$. To
achieve the ELIC design luminosity of 10$^{35}$ cm$^{-2}$s$^{-1}$, R\&D is critical in the
areas of crab crossing, stability of intense ion beams accumulated at stacking, and
electron cooling using a circulator ring. For the former, R\&D is required for the design
of a 1500 MHz multi-cell crab cavity, for understanding the beam dynamics with crab
cavities in both rings, and for achieving phase and amplitude stability requirements.
Understanding beam stability of intense ion beams in boosters and the collider ring also
requires R\&D. One approach is to overcome space charge at injection by increasing the
beam size while preserving the 4D emittance, using a circular painting technique for
stacking similar to the technique proposed at SNS. An alternate approach is to admit a
large beam emittance in the pre-booster and cool it after injection in the collider ring
using stochastic cooling for coasting beam. ELIC's electron cooling concept is unique, in
that it relies on the use of a circulator ring to ease requirements on the average current
from the electron source and on the ERL. Simulation studies are required to establish beam
stability conditions and to optimize the beam and cooling ring operating parameters.
Lastly, the ELIC design requires a dedicated R\&D effort to develop the high-speed data
acquisition and trigger systems that would be needed to accommodate the high collision
frequencies.\\
\section{Theory Opportunities and Initiatives}
\label{Theory}

\subsection{Phenomenology}

The high quality of the RHIC data provides a solid basis for the
quantitative interpretation of the measurements in terms of
fundamental properties of the matter produced in nuclear
collisions. An essential prerequisite for all analyses of this kind is
the sophisticated modeling of the collision dynamics, which must
provide for a detailed description of the evolution of the matter in
space and time.  Such a description would start with the initial
conditions, determined by a detailed quantitative theory of strong
color fields, and would require a theoretical understanding of the
thermalization dynamics leading into the stage of hydrodynamical
expansion.

While relativistic ideal hydrodynamics augmented by hadronic Boltzmann
transport constitutes a solid basis for such modeling efforts, more
sophisticated descriptions involving three-dimensional viscous
relativistic hydrodynamics, as well as detailed simulations of the
propagation of hard probes through the matter and their effect on the
medium will be required to enable quantitative comparisons with the
data. The efforts of a broad community of theorists interested in
interpreting the data in terms of basic material properties, such as
the equation of state, viscosity, stopping power, heavy quark
diffusion constant, and color screening length will increasingly rely
on the availability of sophisticated and validated modeling tools of
this kind.
  
In addition, further progress in extracting quantitative values for
thermodynamic and transport properties of the medium will require the
systematic refinement of the existing treatments of hard probes of hot
and dense matter. Examples of such needs include the next-to-leading
order treatment of radiative parton energy loss, a unified treatment
of elastic and inelastic energy loss mechanisms, and a comprehensive
description of the interaction of heavy quarkonium states with the
medium. These goals are within reach, but will require substantial
investment in theoretical development, discussed below.

\subsection{Lattice QCD}

There are many new opportunities in lattice QCD. These include: a
fully controlled calculation of the equation of state, a better
understanding of the chiral aspects of the finite temperature
transition, detailed study of microscopic properties of QCD matter
such as fluctuations of conserved charges, density correlations,
plasma excitations and transport coefficients. It will become feasible
to map out the phase diagram of QCD at finite temperature and moderate
net baryon density and determine the location of the critical
end-point in the $(T, \mu)$ plane.  This information will be vital for
the success of a future low energy RHIC run, as well as for the
experimental program at GSI/FAIR. Detailed lattice studies of the
temperature region $T_c < T < 3T_c$ will be important benchmarks for
the comparison of data from the RHIC and LHC experiments.

Lattice calculations of the spectral functions are still in their
infancy. To date, almost all such calculations have been done in the
quenched approximation (i.e. neglecting the effect of dynamical
quarks). To have a quantitative impact on RHIC phenomenology such
calculations must be done with dynamical light quarks. This will soon
become feasible due to the expected increase in the computer resources
(the 100 Teraflop Blue Gene supercomputer at BNL and the 1 Petaflop
Blue Gene installation at ANL).

Improved calculations of the meson correlators will also permit
quantitative estimates for some transport coefficients, in particular,
the heavy quark diffusion constant. Up to now, meson correlators have
been studied at zero spatial momentum. In principle, it is
straightforward to extend these calculations to nonzero momenta
$\vec{p}$, where the corresponding spectral functions have a
contribution for energies $\omega < |\vec{p}|$. This component of the
spectral function is related to the scattering of on-shell quarks in
the plasma. Thus lattice calculations may provide for a
non-perturbative insight into the physics of the heavy quark energy
loss. The study of meson correlators at nonzero momentum could also
clarify the dependence of quarkonium suppression on its velocity with
respect to the plasma.

\subsection{Analytical Approaches to Strong Coupling}

In view of the paucity of analytical methods for dynamical problems in
strongly coupled quantum field theories, the value of AdS/CFT
calculations as a tool for gaining qualitative insights is already
well established. Perhaps the semi-quantitative agreement with some
experimental results is a hint that certain properties of strongly
interacting gauge theories are ``universal'' among large classes of
such theories, whereas others are ``microscopic details'', yielding
important differences in vacuum but unimportant in a strongly
interacting quark-gluon plasma which has no quasiparticles. This
question must be addressed by extending AdS/CFT calculations to more
observables and to more, and more QCD-like, gauge theories.

If evidence that the strongly interacting quark-gluon plasmas of QCD
and of theories with a dual string theory description are in the same
universality class accumulates, allowing a better understanding of
what quantities are universal and what quantities are not, the
motivation to address more challenging calculations in strongly
interacting quark-gluon plasmas via AdS/CFT methods will increase.  A
nonzero chemical potential can be added. One can envision implementing
finite volumes of quark-gluon plasma with more and more realistic
geometries, incorporating longitudinal and radial expansion and
elliptic flow. Finally, equilibration can be studied at strong
coupling.

\subsection{New Initiatives}

In order to build and maintain a nuclear theory effort that allows us
to reap the full scientific rewards of the experimental program in
relativistic heavy ion collisions, sound and stable funding for a
broad range of nuclear theory activities of outstanding quality is
needed. In addition to adequate base program support for theorists
addressing questions of fundamental importance for the experimental
program described in Section~\ref{HIFuture}, support for new
initiatives targeting (a) problems of particular programmatic
relevance and requiring the collaboration of theorists at several
institutions, and (b) the rejuvenation of the theory community at the
highest level of excellence, are urgently needed.  Below we describe
specific ideas for such targeted initiatives.

\subsubsection{Programmatic Initiatives}

The recent initiative aimed at providing the hardware needed to
realize the opportunities in lattice QCD thermodynamics must be
continued.  The national lattice initiative demonstrates what can be
accomplished by large, multi-institutional and multifaceted
collaborations with several independent goals but common needs, when
mechanisms and support for their collaborative organization are put
into place.  A similar opportunity has arisen for the theory community
working on phenomenological aspects of relativistic heavy ion
collisions.

The central challenge for the RHIC community now is to progress from
qualitative statements to rigorous quantitative conclusions. The main
obstacle on the path to achieving this goal is the inherently complex
and highly dynamical nature of relativistic heavy-ion
collisions. Quantitative conclusions require sophisticated modeling
and thorough comparison of such models with data. The complexity of
the modeling derives from the fact that reactions traverse two orders
of magnitude of energy density and several distinct phases, each with
different underlying degrees of freedom: a pre-equilibrated phase
characterized by the presence of strong color fields, an approximately
thermalized partonic phase with the characteristics of a nearly ideal
liquid and, finally, a viscous hadronic phase. Experiments provide
three classes of observables: spectra, correlations and fluctuations,
and jets. Each class encompasses a host of hadronic and
electromagnetic species which provide observational access to
different stages of the collision. None of them, taken alone, yields
complete and unambiguous information about any of these stages, but
taken together they hold the promise of fully constraining the
dynamics of the collision and permitting the quantitative extraction
of key properties of the created quark-gluon matter.
 
Doing so will require a full account of the rapid dynamical evolution
of the collision fireball, using sophisticated models which correctly
describe all aspects and stages of its three-dimensional expansion.  A
successful quantitative interpretation of the heavy-ion data will not
be possible without extensive and sophisticated modeling, requiring
close collaboration of the experimental data analysis with the
theoretical modeling effort. Without such an effort, the RHIC physics
program cannot be successfully completed, and the synergies from the
parallel LHC heavy-ion programs cannot be adequately brought to bear
on the physics program of RHIC.  In view of the rapid progress on the
experimental side, the necessary tools for a comprehensive and
quantitative determination of the properties of the medium produced in
relativistic heavy-ion collisions must developed with utmost
urgency. This will require close collaboration between many different
segments of the RHIC theory community, as well as between theory and
experiment. The success of this effort mandates significant additional
investment in theoretical resources in terms of focused collaborative
initiatives.
 
The two established theoretical
milestones\footnote{\footnotesize{2009: ``Perform realistic
three-dimensional numerical simulations to describe the medium and the
conditions required by the collective flow at RHIC.'';\\ \hspace*{5mm}
2010: ``Complete realistic calculations of jet production in a high
density medium for comparison with experiment.''}} in the DOE
performance measures for the RHIC program address limited aspects of
the above challenge. Achievement of these milestones and, more broadly,
realization of the opportunities described in Section~\ref{HIFuture} and
above, are critical to the success of the scientific investments made in
experimental facilities and research.  A collaborative model organized
around common goals like that adopted by the lattice community and in
close coordination with the experimental community may serve many of
these needs.  More focused collaborative structures like topical
centers organized around a specific research program can also be of
value.

Any initiatives of this nature should be launched via a competitive
bidding process, open to the participation of theorists and interested
experimentalists from all universities and national laboratories.
This will ensure that funding of such coordinated efforts targets
phenomenology of the highest quality. It will also ensure that the
theory community as a whole thinks creatively about the most effective
means to accomplish its goals. The size, scientific scope, duration,
degree of geographical localization, and organizational mechanisms of
such initiatives should emerge as outcomes of a competitive process
designed to engage all parts of the theory community.

\subsubsection{Community Oriented Initiatives}

Targeted support in various forms aimed at strengthening the nuclear
theory community by nurturing the careers of creative theorists with
already demonstrated accomplishment, and in this manner attracting the
best theoretical graduate students to work on the rich trove of new
problems which our successes are bringing to light, is critical for
the future of nuclear physics.  The initiatives we describe can easily
be designed for the participation of the entire nuclear physics
community, including all subfields and including both theoretical and
experimental physicists:

\begin{itemize}

\item We recommend the introduction of a national prize fellowship 
program for postdoctoral researchers in nuclear physics.  Winning a
prestigious fellowship in a national competition will raise the
profile of a research career at an early stage and enhance the
visibility of the brightest among our young scientists, and the best
accomplishments of our field, in the larger academic world. Giving the
winners both support and freedom as they launch their research careers
will maximize the scientific impact of these future leaders of the
field at the crucial time when their abilities are fully developed and
their energies are devoted solely to research. Furthermore, the
success and visibility of such a program will have positive impacts on
many additional fronts: it will attract highly talented students to do
graduate work in nuclear physics, retain the best as postdocs working
within our field, raise the visibility of the field by winning the
recognition of the broader physics community that its recipients are
doing outstanding research and continuing onward to successful
careers, and thus assist those seeking to make the case within their
departments or laboratories for hiring of faculty or staff in nuclear
physics.

\item We recommend the introduction of a Nuclear Physics Graduate 
Fellowship, which would identify and support the best graduate
students in the nation who intend to pursue nuclear physics research.
The main objective of this initiative parallels that of the prize
postdoctoral fellowship at one stage earlier, namely to attract the
highest caliber undergraduate students to study nuclear physics.

\item The nuclear physics Outstanding Junior Investigator (OJI) program 
has goals which parallel those of the postdoctoral fellowship, at a
later career stage. This initiative of the DOE should be opened up to
include recently hired staff members in tenure-track positions at the
national laboratories.

\end{itemize}

The base program in nuclear theory must be raised to the point that
outstanding theorists can earn grant support which allows him or her
to build and then {\it maintain} a successful and productive research
effort.  The OJI program and the proposed postdoctoral and graduate
fellowships are part of a concerted effort to further enhance the
excellence of theoretical nuclear physics research in the U.S., but
they will not function as intended without a healthy base program.  If
implemented together with a healthy base program, these initiatives
will yield the kind of breakthrough innovations that can come from
creative research by talented individuals, while at the same time
training people who go on to maximize the effectiveness of more
targeted theoretical pursuits.

\section{Workforce}

The heavy ion community within the broader nuclear physics community
in the United States has been very strong over the last five year
period including the start up and full operational status of the
Relativistic Heavy Ion Collider (RHIC).  The challenge of constructing
a new scale of Nuclear Physics experiments presented significant
questions of labor force, commitment, and coordination amongst the
experimental physics community.  These challenges have been met and
the results are the broad array of high quality precision data from
the four experiments at RHIC (BRAHMS, PHENIX, PHOBOS, and STAR).  Top
young scientists getting their Ph.D's from the RHIC program and
postdoctoral research scientists at the start of the RHIC program are
now new leaders as tenured faculty at our nation's universities and
research scientists at national laboratories.  On the theory side,
again great strides have been made in recruiting top young scientists
and making major contributions in many areas to understand the
experimental data and create a broader picture of the novel state of
nuclear matter under investigation.

Currently the two large experiments (PHENIX and STAR) have over 500
members each (authors in good standing) and with over 100 institutions
in all from around the world.  The author lists have shown steady
growth over the last five years as new institutions (within the United
States and from around the world) have been added and a new graduate
students join the effort.  The smaller experimental groups BRAHMS and
PHOBOS have also been quite successful, and have completed their
programs as of 2006.

As the next phase at RHIC includes not only major detector and
accelerator upgrades at RHIC (including RHIC II luminosity upgrades),
but also the new energy frontier in heavy ion studies at the Large
Hadron Collider (LHC), this labor force will meet new challenges.  In
the last year, both PHENIX and STAR have done a re-assessment of full
time equivalent membership for the next five year period (including a
renewal of memorandum of understandings (MOU) within STAR).  A modest
number of groups will be leaving the RHIC program to focus on the LHC
heavy ion effort and programs elsewhere such as JPARC.  However,
within the United States, few groups are leaving, but rather many will
split their efforts in the future between RHIC and LHC.  Both STAR and
PHENIX project an approximate reduction of 20\% of FTE personel over
the next five year period (2006-2010).  This maintains a strong
program with excellent leadership at RHIC, though will present some
issues for the timely completion of the full detector upgrades and
maintaince of older detector systems.

This 20\% FTE reduction is quite consistent with the increase in FTE
projections from the ALICE, ATLAS, and CMS heavy ion collaborations in
the United States.  ALICE includes 12 US institutions and projects FTE
labor growing from 24 in 2007 to 45 in 2011.  The ALICE construction
project of the Electromagnetic Calorimeter is a major undertaking that
requires significant FTE's over many years.  The ATLAS effort includes
4 US institutions presently and projects 12 FTE growing to 20 FTE in
the next three years.  The CMS effort includes 10 US institutions
presently and projects of order 50 FTE by 2010.  The ATLAS hardware
effort is more targeted with construction limited to Zero Degree
Calorimetry (ZDC), and the CMS effort with ZDC's and also significant
contributions to the high level trigger (HLT).  All groups expect to
make substantial contributions to computing and trigger for heavy ion
specific running.  Although the heavy ion effort at the LHC only has
projected beam time for 1 month each year, the heavy ion groups will
be full members of the LHC experiments that take proton-proton
data for approximately eight months per year.

Thus, the overall heavy ion effort in the United States will remain
strong with a 20\% FTE contingent working at the LHC heavy ion efforts
and a more focussed effort at RHIC.  This allows for a substantial
contribution at the LHC without threatening the existing very strong
program at RHIC. Overall the synergy between the LHC and RHIC projects
will strengthen the heavy ion field and broaden the interests of the
people involved.

\newpage

\section{Education and Outreach}

Education and outreach are central to the missions of the Department
of Energy and the National Science Foundation. They are the
fundamental underpinnings that support the mandates of the agencies to
advance the broad interests of society in academia, medicine,
energy, national security, industry, and government, and to help
ensure United States competitiveness in the physical sciences and
technology.

Similarly, education and outreach are key components of any vision of
the future of the field of nuclear physics. Education is critical to
sustaining a diverse pool of talented nuclear physicists to carry out
a world-leading program of fundamental and applied research, as well
as to train future generations. In addition to these goals, nuclear
science has a long tradition of educating physicists who ultimately
make important contributions to a broad spectrum of societal needs
including medicine, energy, and national security.  This has
most recently been documented in the NSAC Education Report
\cite{NSACEdReport}, for which comprehensive surveys were conducted of
nuclear science PhD degree recipients from 1992-1998. Of that cohort
less than 40\% remained in nuclear science careers in 2003, the
remainder having found rewarding careers in other areas of society.

In order to meet the projected need for nuclear scientists in the
future for basic research and higher education as well as national
needs, the NSAC report recommended that the production of PhDs per
year in nuclear science return to the level of production in the early
1990s, approximately 100 per year. At present the number of nuclear
science PhDs granted per year is approximately 80 and is
decreasing. Continuation of this trend will compromise U.S. leadership
in nuclear science research, resulting in a sub-critical number of
trained researchers, educators, and faculty to meets the nation's
needs.

The collaborations at RHIC (BRAHMS, PHENIX, PHOBOS and STAR) have to
date graduated more than 100 PhD students. These students, who
represent a microcosm reflecting trends in the larger community, have
gone on to postdoctoral fellowships and faculty positions all over the
world, career positions at NNSA laboratories, software companies, and
nuclear energy R\&D; careers in medical physics, teaching, and on Wall
Street. Many other students have received Masters degrees from
work at RHIC, or interned as undergraduates.


In the past, the lack of adequate numbers of U.S. PhDs in nuclear
science has been addressed by recruiting from abroad. However, this
traditional source of talent shows signs of drying up as an increasing
number of attractive opportunities open up in Europe and
Asia. Increasing the number of U.S. citizens who get PhDs in nuclear
science will therefore almost certainly require increased
participation from the full diversity of backgrounds within the
U.S. population. It will also require introducing students to the
concepts of nuclear science and its research before they start
graduate school. These two points are most effectively addressed at
the undergraduate level. Undergraduates are the wellspring of the
pipeline, and the tools and talent exist within the nuclear science
community to make a difference by attacking the problem at this
pressure point. Such an effort best leverages the resources of our
community, building on existing programs (e.g., REU, SULI, CEU, RUI)
and the work of university departments, national laboratories, and
individuals. Therefore, we endorse the first recommendation of the
White Paper: A Vision for Nuclear Science Education and Outreach for
the Next Long Range Plan: {\it The nuclear science community should
increase its involvement and visibility in undergraduate education and
research, so as to increase the number of nuclear science PhDs, and
the number of scientists, engineers and physics teachers exposed to
nuclear science}.

Assuming the success of this initiative, adequate support at the
graduate level for students ultimately attracted to the field is
another key component to insuring the nuclear science workforce of the
future is adequate to the nation's needs.

Outreach to all of nuclear science's stakeholders is also
essential. RHIC has made international headlines since the facility's
commissioning in 1999. The very idea of probing the earliest
microseconds after the Big Bang has sparked people's imaginations in
many directions. RHIC physics is an excellent example of how new and
exciting science can capture public interest if conveyed in an open,
comprehensible way. Conveying this excitement to the public at large
and to teachers and students at all levels is critical for the health
of our field. We applaud Brookhaven National Laboratory's Community,
Education, Government and Public Affairs directorate, which has worked
collaboratively with members of the RHIC community to communicate the
importance and excitement of RHIC science to a diverse community of
stakeholders, skillfully managing perceived negatives (e.g. the
possibility of creating black holes at RHIC) so as to turn potential
controversy into an opportunity for dialogue. This effort serves as an
excellent "best practice" model to the nuclear science community about
how to outreach to its stakeholders: the scientific community, funding
agencies, elected officials, educators and students; the
science-attentive public and general public; the science and
mainstream media. We therefore also endorse the second recommendation
of the White Paper: A Vision for Nuclear Science Education and
Outreach for the Next Long Range Plan: {\it The nuclear science
community should develop and disseminate materials and hands-on
activities that illustrate and demonstrate core nuclear science
principles to a broad array of audiences, so as to enhance public
understanding and appreciation of nuclear science and its value to
society.}

\section{Accelerator R\&D}

The design and construction of new particle accelerators is essential
to the future of nuclear physics.  Increasingly, these accelerators
are quite distinct in character from those planned for high energy
physics.  In addition, compact, low energy accelerators are widely
used in applications of nuclear physics from medicine to cargo
screening.  Nuclear Physics has strongly supported large accelerator
physics efforts at all of its user facilities over many decades and
this will continue to be essential.  However, the support for
accelerator physics and technology at universities has been a very
small ad-hoc effort and not an explicit part of the agencies' program
mission.  It is now time to develop an accelerator science and
technology program that consists of a coordinated effort between the
national laboratories and a modest PI-driven effort at universities
supported by DOE and NSF nuclear physics.  The program should be open
to beam physics research activities relevant to all subfields of
nuclear physics. This effort should be an explicit part of the DOE and
NSF program mission as it would enhance nuclear science in the United
States.

The need for an educational component within this effort was clearly
articulated by the Office of Science within its Occasional Paper
``Accelerator Technology for the Nation'' (2003):

The role of university faculty and students should be expanded in all
aspects of accelerator research from operating accelerators to
advanced accelerator research. This will allow the breadth of
knowledge and expertise that resides at the universities to be brought
to accelerator research, and young scientists will have the
opportunity to learn and become tomorrow's leaders.
	
The program at DOE and NSF would solicit proposals from PI's at
universities to carry out research in accelerator physics relevant to
the priorities of nuclear science.  The proposals would be peer
reviewed and evaluated within the context of the national accelerator
science and technology program.  The grants would support faculty
summer salaries, students, post-docs and equipment.

\newpage

\appendix

\section{Appendix}
\subsection{Program of the Phases of QCD Town Meeting}
\label{Program}

\noindent
Web site is: http://www.physics.rutgers.edu/np/2007lrp-home.html

\vspace{1cm}
\begin{tabular}{||l|l|l||} 
\hline
\multicolumn{3}{||l||}{
\begin{minipage}{\textwidth}\smallskip 
{\bf Friday January 12}\\ (QCD and Hadron Physics Town Meeting meets separately) 
\smallskip\end{minipage}
} 
\\ \hline
9:00-9:05 &  \begin{minipage}{5cm}\smallskip Welcome \smallskip\end{minipage} & \\ \hline
9:05-9:35 & \begin{minipage}{5cm}\smallskip Scientific status of RHIC HI program \smallskip\end{minipage} & P. Steinberg  \\ \hline
9:35-10:05 &  \begin{minipage}{5cm}\smallskip Scientific challenges for the next decade I \smallskip\end{minipage} & U. Wiedemann  \\ \hline
10:05-10:35 &  \begin{minipage}{5cm}\smallskip Scientific challenges for the next decade II  \smallskip\end{minipage} & V. Koch  \\ \hline
10:35-10:55 &  \begin{minipage}{5cm}\smallskip coffee break  \smallskip\end{minipage} &  \\ \hline
10:55-11:20 &  \begin{minipage}{5cm}\smallskip RHIC II: Hard Probes  \smallskip\end{minipage} & A. Drees  \\ \hline
11:20-11:45 &  \begin{minipage}{5cm}\smallskip RHIC II: Low energy  \smallskip\end{minipage} & P. Sorensen  \\ \hline
11:45-12:05 &  \begin{minipage}{5cm}\smallskip ATLAS-US Heavy Ions  \smallskip\end{minipage} & B. Cole  \\ \hline
12:05-12:25 &  \begin{minipage}{5cm}\smallskip ALICE-US  \smallskip\end{minipage} & J. Harris  \\ \hline
12:25-13:30 &  \begin{minipage}{5cm}\smallskip Lunch \smallskip\end{minipage} &  \\ \hline
13:30-13:50 &  \begin{minipage}{5cm}\smallskip CMS-US Heavy Ions  \smallskip\end{minipage} & D. Hofman  \\ \hline
13:50-14:15 &  \begin{minipage}{5cm}\smallskip low-x/eA theory  \smallskip\end{minipage} & R. Venugopalan  \\ \hline
14:15-14:40 &  \begin{minipage}{5cm}\smallskip low-x/eA experiment  \smallskip\end{minipage} & T. Ullrich  \\ \hline
14:40-15:00 &  \begin{minipage}{5cm}\smallskip p/d+A opportunities  \smallskip\end{minipage} & M. Leitch  \\ \hline
15:00-15:20 &  \begin{minipage}{5cm}\smallskip Possible connections to String Theory  \smallskip\end{minipage} & D. Son  \\ \hline
15:20-15:40 &  \begin{minipage}{5cm}\smallskip coffee break \smallskip\end{minipage} &  \\ \hline
15:40-16:00 &  \begin{minipage}{5cm}\smallskip Lattice QCD at Finite Temperature and Density  \smallskip\end{minipage} & F. Karsch  \\ \hline
15:40-16:00 &  \begin{minipage}{5cm}\smallskip Theory Initiatives  \smallskip\end{minipage} & U. Heinz  \\ \hline
16:30-18:30 &  \begin{minipage}{5cm}\smallskip Contributed presentations and discussion  \smallskip\end{minipage} &  \begin{minipage}{5cm}\smallskip P. Petreczky, J. Rafelski, J. Sandweiss \smallskip\end{minipage}  \\ \hline
\hline
\end{tabular}

\vspace{1cm}
\begin{tabular}{||l|l|l||} 
\hline
\multicolumn{3}{||l||}{
\begin{minipage}{\textwidth}\smallskip
{\bf Saturday January 13} \\
(joint session with QCD and Hadron Physics Town Meeting)
\smallskip\end{minipage}
} 
\\ \hline
8:30-8:40 &  \begin{minipage}{5cm}\smallskip Welcome address \smallskip\end{minipage} & Dean David Madigan \\ \hline
8:40-9:25  & \begin{minipage}{5cm}\smallskip JLab 12 GeV upgrade and science program \smallskip\end{minipage}& A. Thomas  \\ \hline
9:25-10:10 & \begin{minipage}{5cm}\smallskip RHIC II upgrade and science program \smallskip\end{minipage}& W. Zajc  \\ \hline
10:10-10:30 & \begin{minipage}{5cm}\smallskip coffee break \smallskip\end{minipage} &  \\ \hline
10:30-10:50 & \begin{minipage}{5cm}\smallskip International opportunities: LHC \smallskip\end{minipage}& B. Wyslouch  \\ \hline
10:50-11:10 & \begin{minipage}{5cm}\smallskip International opportunities: FAIR \smallskip\end{minipage}& W. Henning  \\ \hline
11:10-11:30 & \begin{minipage}{5cm}\smallskip International opportunities: J-PARC \smallskip\end{minipage}& N. Saito  \\ \hline
11:30-11:40 & \begin{minipage}{5cm}\smallskip International opportunities: discussion \smallskip\end{minipage}&  \\ \hline
11:40-12:20 & \begin{minipage}{5cm}\smallskip QCD Theory: challenges, opportunities and community needs \smallskip\end{minipage}& D. Kaplan  \\ \hline
12:20-13:30 & \begin{minipage}{5cm}\smallskip Lunch Break \smallskip\end{minipage} &  \\ \hline
13:30-14:00  & \begin{minipage}{5cm}\smallskip Computational QCD \smallskip\end{minipage}& J. Negele  \\ \hline
14:00-14:35  & \begin{minipage}{5cm}\smallskip Gluons at high density \smallskip\end{minipage}& Y. Kovchegov  \\ \hline
14:35-15:10 & \begin{minipage}{5cm}\smallskip Central questions in nucleon structure \smallskip\end{minipage}& W. Vogelsang  \\ \hline
15:10-15:50 & \begin{minipage}{5cm}\smallskip Opportunities in low-x physics \smallskip\end{minipage}& B. Surrow  \\ \hline
15:50-16:10 & \begin{minipage}{5cm}\smallskip Coffee Break \smallskip\end{minipage} &  \\ \hline
16:10-16:50 & \begin{minipage}{5cm}\smallskip Opportunities in hadron structure \smallskip\end{minipage}& R. Ent  \\ \hline
16:50-17:30 & \begin{minipage}{5cm}\smallskip e+p/A facilities \smallskip\end{minipage}& L. Merminga  \\ \hline
17:30-18:45 & \begin{minipage}{5cm}\smallskip Community input and discussion of priorities \smallskip\end{minipage}&  \\ \hline
\hline
\end{tabular}

\vspace{1cm}
\begin{tabular}{||l|l|l||} 
\hline
\multicolumn{3}{||l||}{
\begin{minipage}{\textwidth}\smallskip
{\bf Sunday January 14}\\ (QCD and Hadron Physics Town Meeting meets separately)
\smallskip\end{minipage}
} 
\\ \hline
9:00-9:20 & \begin{minipage}{5cm}\smallskip American Competitiveness Initiative \smallskip\end{minipage}& E. Hartouni  \\ \hline
9:20-9:40 & \begin{minipage}{5cm}\smallskip Accelerator R\&D at Universities \smallskip\end{minipage}& R. Milner \\ \hline
9:40-10:00 & \begin{minipage}{5cm}\smallskip Status of Theory Support \smallskip\end{minipage}& X.-N. Wang \\ \hline
10:00-10:20 & \begin{minipage}{5cm}\smallskip Computing for experiments \smallskip\end{minipage}& R. Soltz   \\ \hline
10:20-10:40 & \begin{minipage}{5cm}\smallskip Coffee Break \smallskip\end{minipage} &  \\ \hline
10:40-11:00 & \begin{minipage}{5cm}\smallskip Education and Outreach \smallskip\end{minipage}& T. Hallman   \\ \hline
11:00-12:00 & \begin{minipage}{5cm}\smallskip Discussion and White paper planning \smallskip\end{minipage}&  \\ \hline
12:00-12:30 & \begin{minipage}{5cm}\smallskip Lunch Break \smallskip\end{minipage} &  \\ \hline
13:00-14:00 & \begin{minipage}{5cm}\smallskip Discussion and White paper planning \smallskip\end{minipage}&  \\ \hline
\hline
\end{tabular}


\newpage



\begin{thebibliography}{200}

\bibitem{V2Compilation} J. Adams {\it et al.} (STAR), Phys. Rev. {\bf C72} 014904 (2005).

\bibitem{PhenixHF} A. Adare {\it et al.} (PHENIX), nucl-ex/0611018.

\bibitem{PhenixRAA} S. S. Adler  {\it et al.} (PHENIX), Phys. Rev. {\bf C75} 024909 (2007).

\bibitem{STARDihadronHighPt} J. Adams {\it et al.} (STAR), Phys. Rev. Lett. {\bf 91} 072304 (2003).

\bibitem{STARDihadronLowPt} J. Adams {\it et al.} (STAR), Phys. Rev. Lett. {\bf 95} 152301 (2005).

\bibitem{PhenixQM06} B. Sahlmueller {\it et al.} (PHENIX), nucl-ex/0701060.

\bibitem{V2Scale} A.~Adare {\it et al.} (PHENIX), nucl-ex/0608033.

\bibitem{PHOBOSdNdeta} B. B. Back {\it et al.} (PHOBOS), Phys. Rev. Lett. {\bf 91}, 052303 (2003).

\bibitem{Lattice1} C. Bernard {\it et al.}, hep-lat/0611031.

\bibitem{Lattice2} F. Karsch, hep-lat/0701210.

\bibitem{RHICMidtermPlan} Mid-term plan for RHIC: 
\begin{verbatim}
http://www.bnl.gov/HENP/docs/RHICplanning/ RHIC_Mid-termplan_print.pdf
\end{verbatim}

\bibitem{Stephanov} M. A. Stephanov, hep-lat/0701002.

\bibitem{BNLCritPoint} Workshop on ``Can we discover the QCD critical point at
RHIC?'', BNL, March 9-10, 2006
\begin{verbatim}
https://www.bnl.gov/riken/QCDRhic/
\end{verbatim}

\bibitem{eic-wp} A High Luminosity, High Energy Electron-Ion Collider:
\begin{verbatim}
http://www.physics.rutgers.edu/np/070327_EIC_B.pdf
\end{verbatim}

\bibitem{eA-wp} Physics Opportunities with e+A Collisions at an Electron-Ion Collider:
\begin{verbatim}
http://www.phenix.bnl.gov/~dave/eic/PositionPaper_eA.pdf
\end{verbatim}

\bibitem{erhic}eRHIC Zeroth-Order Design Report, Editors: M. Farkhondeh and V. Ptitsyn,
BNL CA-D Note 142, 2004.

\bibitem{elic}Zeroth Order Design Report for the Electron-Ion Collider at CEBAF,
Editors: Ya. Derbenev, L. Merminga, Y. Zhang, April 2007.

\bibitem{NSACEdReport} Education in Nuclear Science: A Status 
Report and Recommendations for the Beginning of the 21st Century, A
Report of the DOE/NSF Nuclear Science Advisory Committee Subcommittee
on Education, November 2004.
\begin{verbatim}
http://www.sc.doe.gov/np/nsac/docs/
NSAC_CR_education_report_final.pdf
\end{verbatim}

\end{thebibliography}
\end{document}